\documentclass[%
 reprint,
superscriptaddress,
 amsmath,amssymb,
 aps,
pra,
]{revtex4-2}

\usepackage{dcolumn}
\usepackage[colorlinks=true, allcolors=blue]{hyperref}
\usepackage[capitalize]{cleveref}
\usepackage{amsfonts}
\usepackage{amssymb}
\usepackage{mathrsfs}
\usepackage{mathtools}
\usepackage{amsmath}
\usepackage{amsthm}
\usepackage{graphicx}
\usepackage{color}
\usepackage{array}
\usepackage[makeroom]{cancel}
\usepackage{changes}
\usepackage{qcircuit}
\usepackage{braket}
\usepackage{bbm}
\usepackage{bm}
\usepackage{comment}
\usepackage{multirow}
\usepackage{cellspace} 
\usepackage{appendix}
\usepackage{tikz}
\usepackage{algorithm}
\usepackage{algpseudocode}
\usetikzlibrary{positioning}
\usetikzlibrary{shapes,arrows,arrows,positioning,fit}
\usetikzlibrary{arrows.meta}
\usetikzlibrary{decorations.pathmorphing}
\usepackage{orcidlink}
\usepackage{leftindex}
\usepackage{empheq}
\usepackage{booktabs}
\usepackage{afterpage}
\usepackage{xcolor}
\usepackage{float}

\crefname{thm}{Theorem}{Theorems}
\crefname{dfn}{Definition}{Definitions}
\crefname{rmk}{Remark}{Remarks}
\crefname{lem}{Lemma}{Lemmas}
\crefname{cor}{Corollary}{Corollaries}
\crefname{section}{Sec.}{Secs.}

\theoremstyle{plain}

\theoremstyle{remark}

\newcommand{\SOC}{\text{SOC}}

\begin{document}


\title{Quantum Algorithms for Photoreactivity in Cancer-Targeted Photosensitizers}

\author{Yanbing Zhou \orcidlink{0000-0001-7673-771X}}
\affiliation{Xanadu, Toronto, ON, M5G 2C8, Canada}

\author{Pablo A. M. Casares \orcidlink{0000-0001-5500-9115}}
\affiliation{Xanadu, Toronto, ON, M5G 2C8, Canada}

\author{Diksha Dhawan \orcidlink{0000-0002-1129-3166}}
\affiliation{Xanadu, Toronto, ON, M5G 2C8, Canada}

\author{Ignacio Loaiza \orcidlink{0000-0001-9630-855X}}
\affiliation{Xanadu, Toronto, ON, M5G 2C8, Canada}

\author{Soran Jahangiri \orcidlink{0000-0002-9988-8841}}
\affiliation{Xanadu, Toronto, ON, M5G 2C8, Canada}

\author{Robert A. Lang \orcidlink{0000-0002-4345-3566}}
\affiliation{Xanadu, Toronto, ON, M5G 2C8, Canada}

\author{Juan Miguel Arrazola \orcidlink{0000-0002-0619-9650}}
\affiliation{Xanadu, Toronto, ON, M5G 2C8, Canada}

\author{Stepan Fomichev \orcidlink{0000-0002-1622-9382}}
\affiliation{Xanadu, Toronto, ON, M5G 2C8, Canada}

\begin{abstract}

Photodynamic therapy (PDT) is a targeted cancer treatment that uses light-activated photosensitizers to generate reactive oxygen species that selectively destroy tumor cells, generally causing less collateral damage than conventional treatments.
However, its clinical success hinges on the availability of photosensitizers with strong optical sensitivity and high efficiency in generating reactive oxygen species. 
While classical computational methods have provided useful insights into photosensitizer design, they struggle to scale and often lack the accuracy needed for these simulations.
In this work, we show how fault-tolerant quantum algorithms can be used to identify promising photosensitizer candidates for PDT. To predict photosensitizer performance, we assess two computational properties. First, we quantify light sensitivity by calculating the cumulative absorption in the therapeutic window with a threshold projection algorithm. Second, we determine the efficiency of reactive oxygen generation by estimating intersystem crossing (ISC) rates using the evolution-proxy approach, complemented by a vibronic dynamic treatment where appropriate.
We apply these algorithms to a clinically relevant and actively pursued class of photosensitizers, BODIPY derivatives, including heavy-atom and transition-metal-substituted systems that are challenging for classical methods.
Our resource estimates, obtained with PennyLane, suggest that systems with active spaces ranging from 11 to 45 spatial orbitals can be simulated using $180$-$350$ logical qubits and Toffoli gate depths between $10^7$ and $10^9$, placing our algorithms within reach of realistic fault-tolerant quantum devices. This paves the way to an efficient quantum-based workflow for designing photosensitizers that can accelerate the discovery of new PDT agents.

\end{abstract}


\maketitle


\section{Introduction}
Broadly acting cancer treatments such as chemotherapy and radiation therapy can harm healthy tissue, motivating the need for more targeted approaches~\cite{SONKIN202418, devita2008, baskar2012}.
Photodynamic therapy (PDT) emerges as a viable alternative and relies on a class of compounds known as photosensitizers, which remain inert until activated by light at specific wavelength ranges matching their absorption bands~\cite{alvarez2024current, el2025advancements}.
Upon light activation, photosensitizer molecules generate reactive oxygen species (ROS) that induce oxidative damage in nearby cells~\cite{agostinis2011}.
Spatial control of illumination confines ROS generation to the targeted region~\cite{macdonald2001basic, Dolmans2003}.

However, existing PDT faces important limitations, particularly in treating deep-seated tumors and achieving high overall therapeutic efficacy \cite{Zehr2025Quantum}. Many existing photosensitizers absorb at wavelengths that are poorly suited for penetrating tissue, limiting their ability to effectively target internal cancers. Even when light delivery is adequate, many photosensitizers exhibit low intersystem crossing efficiency (ISC), limiting the generation of ROS and reducing therapeutic effectiveness~\cite{correia2021, phuong2021}.

\begin{figure*}[t]
    \centering
\includegraphics[width=1.0\textwidth]{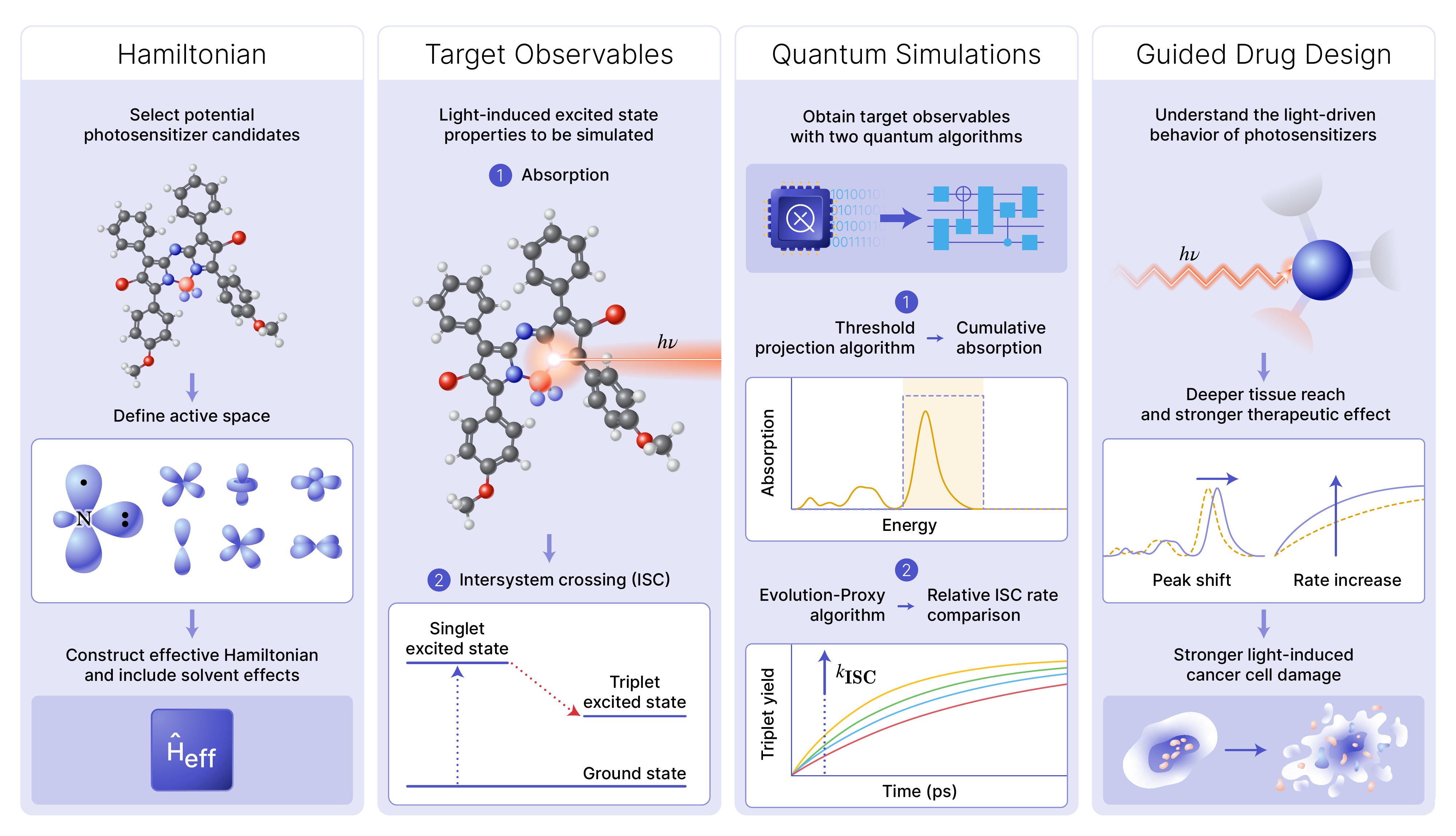} 
    \caption{Workflow for quantum-guided photosensitizer design.
    Left (Hamiltonian): Select a chemically motivated active space and construct an effective Hamiltonian $H_\mathrm{eff}$ using low-rank factorization with solvent embedding.
    Center left (Target observables): Identify two key quantities that determine photosensitizer performance: (i) cumulative absorption within the therapeutic window ($700\text{--}850$~nm), which controls light penetration, and (ii) the intersystem crossing (ISC) rate, which governs population transfer into the reactive triplet manifold.
    Center right (Quantum simulations): Apply two quantum algorithms: (i) threshold projection (\cref{ssec:TP_algo}) to compute cumulative absorption within the therapeutic window; (ii) the evolution–proxy algorithm (\cref{ssec:ISC_evo_proxy_algo}) to rank candidates by their relative ISC rates ($\tilde k_{\text{ISC}}$), estimated from early-time singlet–triplet population transfer.
    Right (Guided drug design): Simulated observables guide molecular edits that red-shift absorption and enhance triplet yield and ISC rate, thereby increasing the efficacy of light-induced cancer therapy.
    }
    \label{fig:hero}
\end{figure*}

Fortunately, the photophysical properties of photosensitizers can be extensively tuned through structural modifications, allowing researchers to significantly boost both the therapeutic absorption window and ISC rates~\cite{correia2021, phuong2021, zhao2021}. However, evaluating the effects of each design choice typically requires extensive experimental synthesis, purification, and characterization work, which are both time- and resource-intensive. An efficient candidate screening procedure would thus be highly desirable. 

\begin{table*}[t]
\centering
\setlength{\tabcolsep}{11pt} 
\renewcommand{\arraystretch}{1.2}
    \begin{tabular}{c c c c c}
        \multicolumn{1}{c}{} & \multicolumn{2}{c}{Cumulative Absorption with threshold projection} & \multicolumn{2}{c}{ISC with evolution proxy}\\ 
        \multicolumn{1}{c}{} & \multicolumn{2}{c}{(Number of Shots $S_\mathrm{dm} = 1.32 \times 10^{2}$)} & \multicolumn{2}{c}{(Number of Shots $S_\mathrm{Had} = 4.83 \times 10^{2}$)}\\
    \cline{2-5}
        $N$ & Qubits & Toffoli per shot & Qubits & Toffoli per shot \\
        \hline
        \hline

    \multicolumn{5}{c}{\centering Original BODIPY} \\ \hline
     $11$ & $177$ & $2.72 \times 10^{7}$ & $176$ & $7.96 \times 10^{7}$ \\
     $15$ & $197$ & $8.44 \times 10^{7}$ & $196$ & $2.43 \times 10^{8}$ \\
     $19$ & $223$ & $1.79 \times 10^{8}$ & $222$ & $5.13 \times 10^{8}$ \\

\hline \multicolumn{5}{c}{\centering Triazolyl aza-BODIPY} \\ \hline 
     $11$ & $177$ & $2.88 \times 10^{7}$ & $176$ & $8.42 \times 10^{7}$ \\
     $19$ & $220$ & $1.24 \times 10^{8}$ & $219$ & $3.55 \times 10^{8}$ \\
     $35$ & $353$ & $9.75 \times 10^{8}$ & $299$ & $3.24 \times 10^{9}$ \\
      
\hline \multicolumn{5}{c}{\centering Brominated aza-BODIPY} \\ \hline
     $17$ & $211$ & $7.64 \times 10^{7}$ & $210$ & $2.20 \times 10^{8}$ \\  
     $21$ & $228$ & $1.63 \times 10^{8}$ & $227$ & $4.67 \times 10^{8}$ \\
     $45$ & $346$ & $2.48 \times 10^{9}$ & $342$ & $5.78 \times 10^{9}$ \\

\hline \multicolumn{5}{c}{\centering Pt-BODIPY
} \\ \hline
     $16$ & $207$ & $7.10 \times 10^{7}$ & $206$ & $2.04 \times 10^{8}$ \\ 
     $24$ & $246$ & $2.91 \times 10^{8}$ & $245$ & $8.31 \times 10^{8}$ \\
     $30$ & $271$ & $6.05 \times 10^{8}$ & $270$ & $1.73 \times 10^{9}$ \\

        \hline
    \end{tabular}
    \caption{Resource estimates, obtained with PennyLane,  for cumulative absorption using the threshold projection algorithm and for ISC rates using the evolution–proxy algorithm, evaluated across active spaces of $N$ spatial orbitals for several BODIPY-class photosensitizer molecules of clinical relevance described in \cref{sec:application}. The trade-off between ancilla qubit count and circuit depth is optimized as described in \cref{ssec:TP_algo}.}
\label{tab:RE_results}
\end{table*}

Computational modeling offers a scalable way to evaluate chemical modifications before synthesis. Classical electronic-structure methods have long provided insight into excited states, ranging from fast empirical and density-functional approaches to more accurate multireference techniques~\cite{Sia_2021, komoto2017thesis, bueno2025, de2018, drzewiecka2021, Berraud-Pache_2020, Momeni_2015}. However, lower-cost approximations often miss the strong electronic correlations that underlie ROS generation~\cite{Momeni_2015}, while higher-accuracy methods scale too steeply to treat the large chromophores used in photosensitizer design~\cite{Krylov2008, Kaliman2017, Pulay2011, Azzi2006, Voggatza2015}.
Even advanced formulations that expand the treatable active-space size have limited reach for full absorption spectra and the many excited states relevant to PDT~\cite{chan2011_DMRG, baiardi2020_DMRG, Wouters_2014, liao2023_DMRG, Dorando_2007, zobel2021, baiardi2019, baiardi2020}.
Quantum computing offers a path to capture strong correlation and many-body excited-state structure, and early demonstrations using variational algorithms have studied vertical excitations and singlet–triplet gaps~\cite{nykanen2024}. Broader VQE-based proposals~\cite{Zehr2025Quantum, ollitrault2020, asthana2023, kim2023, fitzpatrick2024, grimsley2024, chan2021, nader2024_QC, benavides2024, baiardi2021} highlight potential directions, but these methods remain heuristic, lack accuracy guarantees, rely heavily on ansatz design, and do not yet scale to the larger active spaces or full absorption features needed for realistic photosensitizer modeling~\cite{fedorov2021, Tilly_2022, mao2024towards}.

In this work, we develop and apply three fault-tolerant quantum algorithms designed to target two observables directly linked to PDT efficacy: cumulative absorption within the therapeutic window and singlet-triplet intersystem crossing (ISC) rates. 
The first algorithm is a novel threshold projection approach that uses qubitization with low-rank tensor hypercontraction (THC) factorization and spectral filters based on quantum signal processing (QSP) to directly evaluate the cumulative absorption within a specified energy window.
The second is the evolution-proxy algorithm of Ref.~\cite{ODMR_toyota_paper} that enables efficient estimation of relative ISC rates across different systems. 
For completeness, we also include a vibronic dynamics algorithm, based on Ref.~\cite{xanadu_vibronic}, which explicitly incorporates vibronic coupling and captures nonradiative ISC pathways. This more detailed treatment of vibronic effects is presented in \cref{app:ISC_dynamics_algo}.
These algorithms are designed with controllable error sources, so their accuracy is explicitly bounded and systematically improvable.

After describing the key observables of interest and presenting the algorithmic approaches, we apply these methods to a representative set of BODIPY-based photosensitizers relevant to current PDT research, including derivatives containing heavy atoms and transition metals.
For each system, we perform detailed resource estimates of our algorithms using the resource-estimation tools implemented in PennyLane \cite{bergholm2018pennylane}. Our results show that meaningful calculations of key photosensitizer properties beyond the capabilities of classical methods can be achieved with only a few hundred logical qubits and about $10^7$-$10^9$ Toffoli gates, as shown in \cref{tab:RE_results}. This positions the search for advanced photosensitizers as a highly attractive use case for fault-tolerant quantum computers.

The remainder of the paper is organized as follows.
~\cref{sec:background} defines the computational problem and introduces the key physical observables relevant to photosensitizer design---cumulative absorption within the therapeutic window, and the ISC rate---along with a discussion of solvent effects.
\cref{sec:algos} presents the fault-tolerant quantum algorithms we propose for simulating these observables. 
In~\cref{sec:application}, we apply these algorithms to representative photosensitizer systems chosen for their clinical relevance and diversity in electronic structure. 
We describe the selection rationale, simulation setup, and present constant factor resource estimates for each case.
Finally,~\cref{sec:conclusions} contains our conclusions and discusses future directions.

\section{Simulation-guided Photosensitizer Design}
\label{sec:background}
Photosensitizers play a central role in photodynamic therapy, as they absorb light and initiate the sequence of processes responsible for therapeutic action. The therapeutic efficacy of a photosensitizer is governed by a set of photophysical phenomena---absorption of light, relaxation between electronic states, and generation of reactive oxygen species that exhibit cell toxicity. Accurately modeling these processes is critical for rational photosensitizer design but presents substantial computational challenges. In this section, we define the computational problem of photosensitizer design. We lay out the key observables governing performance, explain their physical underpinnings, and discuss the role of solvent effects. Finally, we highlight the limitations of classical methods for accurately modeling these quantities.

\subsection{Fundamentals of photosensitizers}
\label{sec:fundamentals}
The therapeutic efficacy of a photosensitizer depends on two key characteristics. The first is its sensitivity to the light used in therapy. The second is its efficiency at producing the reactive oxygen molecules that kill cancer cells. 

The photosensitizer's sensitivity can be quantified in terms of the molecule's cumulative absorption within the therapeutically relevant near-infrared  window, which for practical PDT applications is often taken to be approximately $700\text{--}850$\,nm. The choice of this window is governed by two considerations: first, the kind of light used must not be inherently harmful to humans. This rules out ultraviolet radiation, which would have been a convenient choice as it aligns much better with typical molecular energy gaps (on the order of a few electronvolts) but is inherently unsafe for humans. 

At the same time, the type of light chosen should also be capable of penetrating into human tissue to treat deep-seated tumors. The two key components making up human tissue, water and hemoglobin, and their respective absorption properties, largely determine the therapeutic window. Water naturally has strong absorption above $900$\,nm, and hemoglobin below about $600\text{--}650$\,nm~\cite{weissleder2001clearer}: from this, the choice of the near-infrared regime of $700\text{--}850$\,nm naturally emerges as the tissue transparency window. An additional advantage of the near-infrared is that in general, light scattering decreases with increasing wavelength, meaning that near-infrared wavelengths can penetrate from several millimeters to several centimeters into tissue, appreciably deeper than visible light. Thus a high-quality photosensitizer must have strong cumulative absorption in this therapeutic window.

Producing the highly reactive ROS is a more complex process. Oxygen molecules are naturally found in their triplet ground state; conversely, most common photosensitizers (and many other organic molecules) are generally found in the singlet ground state. Light absorption alone does not change a molecule's spin state. For the photosensitizer to become effective, it must undergo an internal process that converts the excited singlet state into an excited triplet state. Once in the triplet state, the molecule can relax back to its ground state by transferring energy or an electron to nearby molecular oxygen, producing singlet oxygen in the Type II pathway or superoxide and related radicals in the Type I pathway. This internal transition from a singlet to a triplet state is known as intersystem crossing (ISC), and its efficiency determines how much population reaches the long-lived triplet states. For many photosensitizer families, a high ISC rate is therefore a desirable property, as it directly boosts the formation of ROS.

We can therefore translate the photosensitizer's therapeutic goals---light sensitivity and efficient reactive oxygen generation---into two quantum-mechanical computational variables: cumulative absorption in the therapeutic window and the ISC rate. Most conventional photosensitizers exhibit poor characteristics in these two respects: they naturally absorb in the visible range rather than in the near-infrared; and their spin-orbit coupling, and thus ISC rates, are weak.
However, structural modification of photosensitizer molecules holds great potential for boosting their therapeutic window absorption and ISC rates. This is where efficient simulation of these two properties could help guide the development of more efficient photosensitizers by screening viable candidates with respect to these two key properties. 

\subsection{Key Observables}

To simulate and optimize these two key properties---cumulative absorption in the therapeutic window and ISC rate---at the quantum level, we must model the key excitation and relaxation pathways in the photosensitizer. These begin with photon absorption, followed by intersystem crossing transitions, and end with reactive oxygen generation. We now review this mechanism in detail to precisely define the observables targeted by our quantum algorithms.

\begin{figure}[t]
    \centering
    \begin{tikzpicture}[scale=0.9]

    \node at (-1.7,2.0) {\scriptsize E};
    \draw[very thick, ->] (-1.5,-0.2) -- (-1.5, 3.1);

    \node at (1.9,-0.4) {\small PS};
    
    \draw[very thick] (0,0) -- (4.2,0); 
    \node at (-0.3, -0.4) {\scriptsize \shortstack{ground states \\ $n=0$}};
    \draw[very thick] (0,2.7) -- (1.5,2.7) ;
    \node at (-0.3, 3.5){\scriptsize \shortstack{singlet \\ excited states \\ $n=1,...; S=0$}};
    \node at (0.7,2.8) {\scriptsize $\dots$};
    \draw[very thick] (0,2.9) -- (1.5,2.9) 
    ;

    \draw[very thick] (2.7,1.8) -- (4.2,1.8) ; 
    \node at (4.2, 2.6) {\scriptsize \shortstack{triplet \\ excited states \\ $n=1,...; S=1$}};

    \draw[decorate, decoration={snake, amplitude=0.4mm}, ->, thick] 
      (-1.2,1) -- (0.1,1) node[midway, above] {\scriptsize $h\nu$};

    \draw[->, thick] (0.35, 0) -- (0.35, 2.85);

    \draw[->, thick] (0.7, 2.85) --(0.7, 0)  ;

    \draw[dashed, ultra thick, ->] (1.5,2.7) -- (2.7,1.85) node[midway, above right] {\small ISC};

    \draw[decorate,decoration={snake, amplitude=0.4mm},->, thick] (1.3,2.7) -- (1.3,0.05) node[midway, right] {\tiny fluores.};

    \draw[decorate,decoration={snake, amplitude=0.4mm},->, thick] (4.0,1.8) -- (3.2,0.05) node[midway, left] {\tiny phosphores.};

    \draw[->, very thick] (4.4, 1.55) .. controls (4.9,1.3) and (4.9,0.5) .. (4.4,0.1);
    \draw[->, very thick] (5.5, 0.1) .. controls (5.0,0.25) and (5.0,1.15) .. (5.5, 1.55);

    \draw[->, very thick] (4.4, 1.7) -- (5.6, 2.2) ;
    \node at (6.9, 2.9) {\scriptsize \shortstack{Type I \\ electron transfer}};
    \node at (5.9, 2.3) {\scriptsize \shortstack{$^{.}O_2^-$}};
    \node at (5.3, 1.9) {\scriptsize \shortstack{$e^-$}};

    \node at (5.9, 0.0) {\footnotesize $^3\text{O}_2$};
    \node at (5.9, 1.6) {\footnotesize $^1\text{O}_2$};

    \node[draw, circle, inner sep=1pt, font=\scriptsize] at (-0.6, 0.6) {1};
    \node[draw, circle, inner sep=1pt, font=\scriptsize] at (1.9, 2.) {2};    
    \node[draw, circle, inner sep=1pt, font=\scriptsize] at (5.4, 0.8) {3}; 
    \node at (6.7, 0.8) {\scriptsize \shortstack{Type II \\ energy transfer}};
    \node[draw, circle, inner sep=1pt, font=\scriptsize] at (4.9, 1.65) {4};   
    \node[draw, circle, inner sep=1pt, font=\scriptsize] at (0.93, 0.9) {5};  

    \end{tikzpicture}
\caption{Photophysical processes underlying the PDT application studied in this work. 
\textcircled{1} Upon absorbing a photon $h\nu$, the photosensitizer is promoted from its singlet ground state (with quantum numbers $n = 0; \ S = 0$) to excited singlet states (with $n = 1,...; \ S = 0$). 
\textcircled{2} The excited-state population may undergo intersystem crossing, a spin-forbidden transition, to the triplet manifold (with $S = 1$), typically to the lowest triplet state. 
\textcircled{3} Photosensitizer in its triplet excited state can subsequently transfer energy to ground state oxygen $^3\text{O}_2$, converting it into singlet oxygen $^1\text{O}_2$, a reactive oxygen species (ROS) that drives local oxidative damage in cells. 
\textcircled{4} Excited photosensitizer triggers an electron transfer to nearby oxygen, forming superoxide radicals $^{.}O_2^{-}$ as Type I ROS.
\textcircled{5} Vibrational relaxation occurs as a fast, nonradiative process within a given electronic manifold, dissipating excess vibrational energy and allowing the system to settle into its vibrational ground level before undergoing fluorescence, phosphorescence, or ISC.
}

    \label{fig:type2_PDT_diagram}
\end{figure}
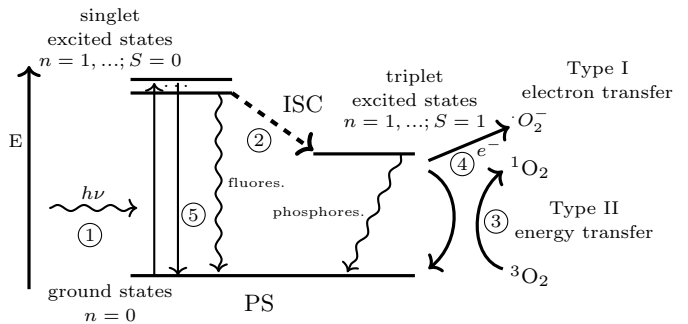

The light activation process typically begins with photosensitizer molecules in the electronic ground state, denoted $\ket{E_{n=0; S=0}}$. 
Here, $n$ labels the electronic excitation level, with $n=0$ denoting the ground state, $n=1$ the first excited state, and so on; $S$ denotes the total spin quantum number of the electronic state. States with $S=0$ are singlets, having paired electron spins, while those with $S=1$ are triplets, with two unpaired electrons aligned in parallel.
Upon irradiation with external light (step~\textcircled{1} in \cref{fig:type2_PDT_diagram}), a photosensitizer absorbs a photon and transitions to an excited singlet state $\ket{E_{n=1; S=0}}$ or higher.

After light absorption and excitation from the singlet ground to a singlet excited state, effective therapeutic action requires the photosensitizer to undergo intersystem crossing---a spin-forbidden and non-radiative transition---to a triplet excited state $\ket{E_{n; S=1; M}}$ (step~\textcircled{2} in \cref{fig:type2_PDT_diagram}). The variable $M \in \{-1, 0, +1\}$ denotes the spin projection quantum number. 
The ISC process is driven by the spin–orbit coupling (SOC) present in the system, and can potentially also be affected by vibronic coupling between electronic and vibrational eigenstates. 
Given that the ground state is assumed to be a singlet, the triplet state $\ket{E_{1; S=1; M}}$ is typically long-lived, as de-excitation requires a spin flip. This enables the photosensitizer to subsequently interact with ambient molecular oxygen $^3O_2$ (step~\textcircled{3} and \textcircled{4} in \cref{fig:type2_PDT_diagram}), which exists in a triplet ground state.

Ground-state triplet oxygen ($^3O_2$) is the naturally abundant form of oxygen~\cite{martemucci2022, wilson2008}.
Photosensitizers in their triplet state can react through two distinctive pathways. In the Type II pathway, the triplet photosensitizer transfers energy to $^3O_2$, generating singlet oxygen $^1O_2$, an electronically excited and highly reactive form of oxygen (step~\textcircled{3} in \cref{fig:type2_PDT_diagram}). 
In parallel, many photosensitizers also engage in Type I reactivity (step~\textcircled{4}), where the triplet state donates an electron or hydrogen atom to nearby acceptors, including $\mathrm{O}_2$, producing superoxide and related radicals. Type I pathways becomes particularly important under hypoxic tumor conditions, where limited oxygen availability suppresses efficient singlet-oxygen production~\cite{CASTANO20051, KWIATKOWSKI20181098}. 
Singlet oxygen and Type I radicals both induce oxidative stress, damaging key biomolecules such as lipids, proteins, and DNA, ultimately triggering the death of the cancerous cell~\cite{agnes2012, an2024oxidative}. 
Though steps~\textcircled{3} and ~\textcircled{4} are crucial for therapeutic action, we do not simulate these processes directly. 
Unlike absorption and intersystem crossing, which are intrinsic molecular properties of the photosensitizer, simulating ROS generation requires treating the coupled photosensitizer-acceptor system and depends strongly on the biological environment.
This makes the process highly system-specific and beyond the scope of a general photosensitizer simulation framework.

Instead, we focus on the preceding photophysical stages (step~\textcircled{1} and \textcircled{2}), which govern the yield of reactive photosensitizer triplet states. 
These upstream properties serve as the primary computational targets as they are intrinsic to the molecule and, in many cases of practical interest, correlate with downstream ROS generation.
These simulations help rank candidates before complex biological modeling becomes relevant, allowing us to assess intrinsic photosensitizer performance in a transferable, system-independent manner.

Having introduced the photosensitizer excitation process in \cref{fig:type2_PDT_diagram}, we can now define precisely the two computational variables our algorithms will target.

\textbf{Cumulative Absorption:} 
Accurate prediction of a system's optical response requires knowledge of both the excitation energies  
$\Delta E_f \!=\! E_f-E_0$ and corresponding dipole transition strengths  
$|\langle E_f|\hat D|E_0\rangle|^{2}$, where $E_0$ is the ground-state energy, $E_f$ denotes the energy of an excited state, and $\hat{D}$ is the electric dipole operator. The one-photon absorption cross section has the form 
\begin{equation}
\sigma_A(\omega) = \frac{4\pi} {3\hbar c}\omega\sum_{f} \frac{\eta |\bra{E_f}\hat D\ket{E_0}|^2}{(E_f - E_0)^2 + \eta^2},
\end{equation}
where $\omega$ is the angular frequency of the incident light and $\eta$ is a Lorentzian broadening. As discussed in \cref{sec:fundamentals}, we focus on the near-infrared range  
$[700,850]~\,\text{nm}$, corresponding to angular-frequency bounds  
$\omega_{\text{lo}}$ and $\omega_{\text{hi}}$.  
The cumulative absorption within this spectral window is defined as
\begin{equation}
    A_{\text{window}}
   \;=\;
\int_{\omega_{\text{lo}}}^{\omega_{\text{hi}}}
   d\omega\,\sigma_A(\omega) .
\label{eq:cumulative_absorption_integral}
\end{equation}
In the stick-spectrum limit $\eta\!\to\!0$, each Lorentzian reduces to a delta function, and Eq.~\eqref{eq:cumulative_absorption_integral} simplifies to a dipole-weighted discrete sum over the final states whose excitation energies fall within the desired energy window:
\begin{equation}
A_{\text{window}}
   \;=\;
   \sum_{\Delta E_f\in[E_{\text{lo}},E_{\text{hi}}]}
        |\langle E_f|\hat D|E_0\rangle|^{2}.
\label{eq:cumulative_absorption_sum}
\end{equation}

\textbf{Intersystem Crossing:} ISC is a non-radiative transition between electronic states of different spin multiplicities---typically from an excited singlet to a triplet state---and is formally spin-forbidden. However, spin–orbit coupling (SOC) enables this process by mixing spin and orbital angular momentum. The efficiency of ISC is thus strongly modulated by the presence of heavy atoms, such as bromine or transition-metal centers, which enhance SOC and promote population transfer to the triplet manifold \cite{Kamkaew2013BODIPY}.

To quantify ISC efficiency, we compute the SOC matrix elements between singlet and triplet states, which determine the rates of spin-flip transitions. 
These couplings are derived from the one-electron part of the Breit–Pauli spin–orbit Hamiltonian, denoted $H_{\text{SOC}}$.
The two-electron terms can, in principle, be included in a mean-field approximation to retain a one-body form~\cite{neese2005efficient}. However, since such a contribution is generally small compared to the one-electron term, we follow prior work in neglecting it~\cite{ODMR_toyota_paper}. 
The SOC operator originates from the relativistic interaction between the spin and orbital motion of electrons in the electrostatic field of the nuclei. In first quantization, this operator takes the form:
\begin{equation}
H_\text{SOC} = \frac{\alpha^2}{2} \sum_{i=1}^{N_e} \sum_{I=1}^{N_\text{A}} \frac{Z_I}{|\bm{r}_i-\bm{R}_I|^3} \left[ (\bm{r}_i - \bm{R}_I) \times \bm{p}_i \right] \cdot \bm{s}_i,
\end{equation}
where $N_e$ is the number of electrons and $N_\text{A}$ is the number of atoms, $\bm{r}_i$ is the position operator of electron $i$, $\bm{R}_I$ and $Z_I$ are the position and atomic number of nucleus $I$, $\bm{p}_i$ and $\bm{s}_i$ are the momentum and spin operators for electron $i$. 
This expression is then integrated over basis functions to yield one-electron spin–orbit integrals $h^{\text{soc}}_{p\sigma, q\tau}$ in the atomic orbital basis, which are transformed to the molecular orbital basis and expressed in second quantization as:
\begin{equation}
H_{\text{SOC}} = \sum_{p,q=1}^N \sum_{\sigma,\tau} h^{\text{soc}}_{p\sigma,q\tau}~c^\dagger_{p\sigma} c_{q\tau}.
\end{equation}
Here, $p, q$ index spatial orbitals and $\sigma, \tau$ label spin projections. The matrix elements $h^{\text{soc}}_{p\sigma,q\tau}$ encode the atomic and spin-angular momentum structure of the SOC interaction.
Following the spin-tensor formalism described in Ref.~\cite{ODMR_toyota_paper}, we further decompose $H_{\text{SOC}}$ into components that transform as spin tensor operators $T^{S,M}$ under SU(2) symmetry:
\begin{equation}
H_\text{SOC} = H_\text{SOC}^{0,0} + H_\text{SOC}^{1,0} + H_\text{SOC}^{1,1} + H_\text{SOC}^{1,-1},
\end{equation}
where each $H_\text{SOC}^{S,M}$ contains a well-defined pattern of spin selection rules. In particular, $H_\text{SOC}^{1,0}$ preserves spin projection ($\Delta M = 0$) but can change the total spin ($\Delta S = \pm 1$), while $H_\text{SOC}^{1,\pm1}$ induces transitions with $\Delta M = \pm1$. This decomposition enables us to isolate different intersystem crossing channels based on spin symmetry.

Larger couplings correlate with more efficient triplet population production, which in turn promotes stronger ROS-generating capacity. 
In the weak coupling regime, where the perturbation introduced by $\hat{H}_{\text{SOC}}$ is small relative to the adiabatic energy gap between spin manifolds, the intersystem crossing rate $k_{\text{ISC}}$ is governed by Fermi’s golden rule~\cite{royea_fajardo_lewis_1997, Marian_2012, penfold2018}:
\begin{equation}
k_{\text{ISC}} = \frac{2\pi}{\hbar} \sum_f \left| \langle \Psi_f | \hat{H}_{\text{SOC}} | \Psi_i \rangle \right|^2 \delta(E_f - E_i),
\end{equation}
where $\Psi_i$ and $\Psi_f$ denote the total molecular wavefunctions (including both electronic and vibrational components) of the initial singlet and final triplet states, and $\delta$ enforces energy conservation.
When the spin–orbit interaction is separable from nuclear motion (i.e., assuming direct SOC), the total matrix element factorizes, and the expression simplifies to~\cite{penfold2018}:
\begin{equation}
k_{\text{ISC}} = \frac{2\pi}{\hbar} \sum_{f} \left| \langle E_f | \hat{H}_{\text{SOC}} | E_i \rangle \right|^2 \sum_k \left| \langle \nu_{fk} | \nu_{ia} \rangle \right|^2 \delta(E_{ia} - E_{fk}),
\end{equation}
where $\ket{E}$ are the electronic wavefunctions, $\nu$ the nuclear vibrational states, and the second summation quantifies vibrational density-of-states overlap.
We will take advantage of this electron-nuclear separability in our first algorithm for simulating ISC (\cref{ssec:ISC_evo_proxy_algo}), where we will assume that the dominant contribution to ISC arises from the electronic SOC operator, while its dependence on nuclear displacements can be neglected. The more general situation, where spin–vibronic coupling modulates the transition, is treated separately in our explicit spin–vibronic Hamiltonian framework (\cref{app:ISC_dynamics_algo}).
The squared matrix element $\left| \langle E_f | \hat{H}_{\text{SOC}} | E_i \rangle \right|^2$ gives the probability of the spin-flip transition, and the sum runs over all energetically accessible triplet states.
To efficiently capture the relevant spin-flip transitions, we decompose the SOC operator into its irreducible tensor components $ H_{\mathrm{SOC}}^{1,M} $. Our analysis focuses on the $E_{n=1; S=0} \rightarrow  E_{n=1; S=1; M}$ channel, which is both energetically accessible and thermodynamically favored under physiological conditions~\cite{przygoda2023, yu2022, ponte2022}.

Together, these observables define well-motivated computational targets 
that reduce the problem of discovering improved photosensitizers to screening candidate molecules for strong cumulative absorption in the therapeutic window and high ISC rates.

\subsection{Solvent Effect}

To properly simulate the key observables discussed above, it is crucial to account for the influence of the surrounding environment. 
In biological environments like blood plasma and intracellular fluid, \textit{solvent effects} play a non-negligible role in shaping the photophysical behavior of photosensitizers.
These effects can shift excitation energies, modify transition dipole moments, and influence SOC, all of which affect absorption spectra and ISC rates~\cite{Telegin2021, Kalyagin2022, Laine2016, Pothoczki2021, Marian2012}.
Ignoring solvent-induced effects risks errors in modeling photosensitizer performance under biological conditions.

Water is the primary component of biological fluids and thus the main solvent of relevance for modeling photosensitizer behavior in physiological environments. Its high dielectric constant stabilizes polar or charge-separated excited states, modifying excitation energies, transition dipole strengths, and singlet–triplet energy gaps---key quantities that affect both absorption and intersystem crossing efficiency.

Solvent effects in molecular simulations are typically treated using implicit or explicit models, depending on whether the system is in a so-called resonant or non-resonant regime. This distinction reflects how dynamically the solvent responds to solute excitation. In the resonant regime, where the solute excitation energy overlaps with solvent absorption bands, strong coupling necessitates time-dependent modeling of solvent polarization. In the non-resonant regime, where the excitation energy lies well outside the solvent's absorption spectrum, the solvent acts as a passive dielectric with negligible dynamical feedback~\cite{wildman_2019, liu_2019}. 

In PDT, we seek photosensitizers with strong absorption in the $700\text{--}850$\,nm range, where water shows minimal electronic or vibrational absorption. 
This places the system in the non-resonant regime, justifying the use of a frequency-independent polarizable continuum model (PCM) to treat solvent effects as static corrections to the solute Hamiltonian without modeling solvent dynamics explicitly~\cite{PCM}. 
While solvent-induced shifts can influence the position of absorption features, these effects are captured through static polarization in a frequency-independent PCM. This approach remains appropriate for our setting and avoids the added complexity of modeling dynamic solvent response, which is only necessary when solvent resonances directly overlap with electronic transitions.
This choice of frequency-independent model is practical for both classical and quantum simulations~\cite{Gotardo2017, drzewiecka2021}: solvent polarization is incorporated through a classical self-consistent PCM treatment, after which the resulting Hamiltonian remains fixed. This avoids quantum–classical iterations and enables precomputed static Hamiltonians for quantum algorithms, reducing circuit complexity and complying with current hardware constraints (see Section~\ref{ssec:bdp_active_space} for further discussion).

While water remains the solvent of primary interest in physiological contexts, other environments may involve solvents with absorption features overlapping the excitation spectrum. 
In such cases, the assumption of frequency-independent dielectric screening breaks down. In practice, a PCM-type continuum model serves as the baseline description, but more structured microenvironments, such as lipid membranes or nucleic-acid binding pockets, can require a treatment that captures solvent response beyond static dielectric screening. 
To accommodate frequency-dependent solvent response, we outline an alternative quantum embedding scheme (Appendix~\ref{app:bosonic_solvent}) in which polarization is modeled using a discrete set of harmonic oscillators---bosonic modes---fit to the solvent’s dielectric function. 
This construction captures dynamic screening via the boson spectral density and enables treatment of resonant solvent effects when the bosonic modes are explicitly evolved.

In the present non-resonant setting (e.g., water in the NIR), the bosons remain frozen in their ground state, and integrating them out yields an energy-filtered, static correction to the one-electron Hamiltonian of the solute. This provides a static reaction-field correction as a frequency-independent PCM.
This framework, however, provides a natural path to include frequency-dependent (resonant) solvent response when needed.

\subsection{Limitations of classical methods}

Accurate modeling of absorption spectra and ISC rates relies on precise calculation of photosensitizer excited states. 
The most widely used method, time-dependent density functional theory (TD-DFT), often fails to deliver quantitative, and in some cases even qualitative, accuracy due to its intrinsic limitations.
As a single-reference method, TD-DFT struggles with systems that exhibit strong correlation or near-degeneracies, which are common features in photosensitizers incorporating heavy atoms, transition metals, or extended conjugation. Moreover, TD-DFT results are highly sensitive to the choice of exchange-correlation functional, introducing empirical ambiguity and limiting its predictive power across diverse chemical systems. As a concrete example, these issues lead to systematic errors of $0.3\text{--}0.6$\,eV in computed vertical excitation energies of a series of BODIPY-based photosensitizers~\cite{Momeni_2015, Sia_2021}, which translate to wavelength shifts of $180\text{--}600$\,nm. Considering the width of the therapeutic window is only $150$\,nm, such deviations can easily shift predicted absorption peaks outside the window, significantly undermining the accuracy of screening and design workflows. 

Post-DFT, wavefunction-based methods, such as equation-of-motion coupled cluster with singles and doubles (EOM-CCSD), and multireference approaches like complete active space second-order perturbation theory (CASPT2), can improve upon TD-DFT by more rigorously capturing electron correlation. 
For many systems, these methods can reduce excitation energy errors to below 0.2\,eV, and in certain situations might even achieve chemical accuracy for moderately sized systems.
However, the general applicability of such methods is constrained by steep computational scaling and large memory footprints. More specifically, the EOM-CCSD method exhibits a $\mathcal{O}(N^6)$ scaling due to its underlying tensor contractions without a guarantee of achieving the required accuracy
~\cite{Krylov2008, Kaliman2017}. 
For CASPT2 and related multireference perturbation methods, the active space is typically limited to approximately 16 orbitals~\cite{Pulay2011, Azzi2006, Voggatza2015}, as the number of determinants in the underlying CASSCF wavefunction grows factorially with the number of electrons and orbitals.

DMRG provides a powerful alternative to conventional CAS-based methods, enabling the treatment of much larger active spaces. 
However, while ground-state studies have reached hundreds of orbitals, excited-state spectroscopy applications remain far more limited—particularly when both a large active space and an extensive excited-state manifold are required~\cite{baiardi2020}.
The need to describe many singlet and triplet states simultaneously, combined with unstable state-averaged optimizations and the cost of dynamic correlation corrections, has kept practical spectroscopy studies confined to modest active spaces and a limited number of excited states~\cite{zobel2021, phung2021, Ghosh_2008}.

The recent review by Zehr et al.~\cite{Zehr2025Quantum} echoed these concerns in the context of photosensitizer design. While acknowledging that techniques like DMRG and QMC can offer high accuracy for systems with strong correlation, the authors emphasize that the complexity of realistic photosensitizers continues to outstrip the capability of available classical methods. These challenges, especially in the context of excited-state modeling across molecules, underscore the need for alternative computational strategies that can achieve high accuracy together with improved scalability.

\section{Quantum Algorithms}
\label{sec:algos}

Quantum computing offers a natural framework for simulating excited-state-dependent observables such as absorption spectra and intersystem crossing rates, as these calculations can typically be efficiently expressed in terms of time evolution of many-body quantum states. 
Unlike systematically improvable classical algorithms like CAS that suffer from exponential scaling in the system size, quantum algorithms can represent and coherently evolve entangled many-body wavefunctions with resources that scale polynomially in the number of orbitals and electrons.
This offers the potential to simulate strongly correlated systems with higher accuracy and scalability as fault-tolerant quantum hardware matures.
In the following sections \ref{ssec:TP_algo} and \ref{ssec:ISC_evo_proxy_algo}, we present an end-to-end computational pipeline for estimating cumulative therapeutic window absorption and ISC rates. For completeness, we also provide a high-level description and order-of-magnitude resource estimates for a vibronic dynamics–based treatment of ISC (\cref{app:ISC_dynamics_algo}) for systems with strong vibronic coupling.

\subsection{Threshold projection with quantum signal processing (QSP) for cumulative absorption} \label{ssec:TP_algo}

When it comes to maximizing photosensitizer sensitivity, the key observable of interest is the \textit{cumulative absorption} within the therapeutic window ($700\text{--}850$\,nm), as defined in Eq.~\ref{eq:cumulative_absorption_sum} of \cref{sec:background}. This quantity measures the total dipole-weighted transition probability from the ground state to excited states within that window.
Thus the goal is to compute the integrated intensity over an energy range, rather than resolving precisely the fine structure of the absorption spectrum---as might be needed in a spectral fingerprinting application \cite{fomichev2024,loaiza2025simulating}. 

Taking advantage of this fact, we propose a quantum algorithm to obtain the cumulative absorption in a fixed energy window. The approach first prepares an initial state encoding the system's dipole transition amplitudes. Next, it synthesizes a projector onto the energy window of interest. The projector can, in principle, be implemented with quantum phase estimation combined with the median lemma to control the failure probability~\cite{nagaj2009fast}. In this work, however, we adopt an approach leveraging qubitization together with generalized quantum signal processing~\cite{motlagh2024generalized}, which directly realizes a polynomial approximation to the desired threshold function and achieves sharp filtering with lower circuit depth. 

The cumulative absorption is formally given by the squared overlap of the dipole-excited state with its projection into the target energy window. Operationally, this overlap is estimated by sampling the ancilla register of the projection circuit: the average ancilla outcome corresponds to the fraction of dipole-weighted population inside the window.
The key steps of the algorithm are detailed below.

\textbf{State preparation:} 
The quantum algorithm begins with preparation of the initial state 
\begin{equation}
|\psi\rangle = \hat{D} |E_0\rangle = \sum_f \langle E_f | \hat{D} | E_0 \rangle\, |E_f\rangle,
\end{equation}
where $\{|E_f\rangle\}$ are the excited eigenstates of the Hamiltonian, with $\ket{E_0}$ the ground state. 
The operator $\hat{D}$ is not unitary, so $|\psi\rangle$ must be normalised before preparation in the quantum computer. Defining the total dipole transition strength
\begin{equation}\label{eq:normalization_Nd}
\mathcal{N}_D \;=\; |\langle \psi | \psi \rangle| ^2
              \;=\; \sum_f |\langle E_f | \hat{D} | E_0 \rangle|^2,
\end{equation}
we prepare the normalised state
\begin{equation}
|\widetilde{\psi}\rangle
   \;=\; \frac{|\psi\rangle}{\sqrt{\mathcal{N}_D}}
   \;=\; \sum_f
          \frac{\langle E_f | \hat{D} | E_0 \rangle}
               {\sqrt{\mathcal{N}_D}}\;
          |E_f\rangle .
\label{eq:normalized_state}
\end{equation}
Our algorithms assume this state is classically evaluated to sufficiently high accuracy. This is a reasonable assumption, given the well-known capability of classical methods such as DMRG or selective configuration interaction for evaluating the ground state of molecular Hamiltonians for the kind of system sizes considered in our application. At the same time, spectral properties of these systems remain well outside the capabilities of such methods.
Next, we prepare this state in the quantum computer. Because the amplitudes in \cref{eq:normalized_state} are known classically, this step reduces to preparing a sparse, classically specified state. We use the sum-of-Slaters method~\cite{Fomichev2024_SoS}, which prepares such states with depth $\mathcal{O}(D \log D)$, where $D$ is the number of nonzero components. The method relies on quantum read-only memory (QROM) to load the amplitudes coherently and keeps the overall overhead modest.

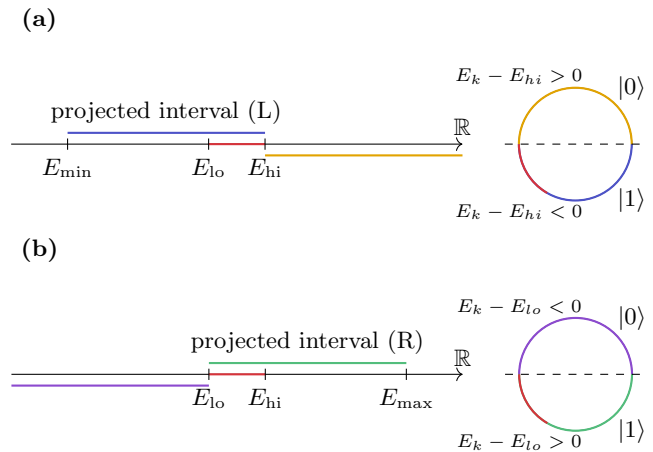
\begin{figure}
\definecolor{XanaduBlue}{HTML}{4D53C8}
\definecolor{XanaduRed}{HTML}{D7333B}
\definecolor{XanaduGreen}{HTML}{52BD7C}
\definecolor{XanaduOrange}{HTML}{E2A300}
\definecolor{XanaduPurple}{HTML}{8C4ACF}
\centering
\begin{tikzpicture}[scale=0.75]
\node at (-5.5,2.2) {\textbf{(a)}};
\draw[->] (-6,0) -- (2,0) node[above] {$\mathbb{R}$};
\draw[thick,XanaduBlue] (-5,0.2) -- (-1.5,0.2) node[midway, above, text=black] {projected interval (L)};
\draw[thick,XanaduOrange] (-1.5,-0.2) -- (2,-0.2);
\draw[thick,XanaduRed] (-2.5,0) -- (-1.5,0);
\foreach \x/\l in {-5/$E_\text{min}$, -2.5/$E_\text{lo}$,-1.5/$E_\text{hi}$}
    \draw (\x,0.1) -- (\x,-0.1) node[below] {\l};
\draw (4,0) circle (1cm);
\draw[thick,XanaduBlue] (5,0) arc (0:-180:1cm);
\draw[thick,XanaduRed] (3,0) arc (180:240:1cm);
\draw[thick,XanaduOrange] (3,0) arc (180:0:1cm);
\node at (5,1) {$\ket{0}$};
\node at (5,-1) {$\ket{1}$};
\draw[dashed] (2.75,0) -- (5.25,0);
\node at (3.0, 1.2) {\scriptsize \shortstack{$E_k-E_{hi} > 0$}};
\node at (3.0, -1.2) {\scriptsize \shortstack{$E_k-E_{hi} < 0$}};

\end{tikzpicture}
\vspace{1.5em}
\begin{tikzpicture}[scale=0.75]
\node at (-5.5,2.2) {\textbf{(b)}};
\draw[->] (-6,0) -- (2,0) node[above] {$\mathbb{R}$};
\draw[thick,XanaduPurple] (-6,-0.2) -- (-2.5,-0.2);
\draw[thick,XanaduGreen] (-2.5,0.2) -- (1,0.2) node[midway, above, text=black] {projected interval (R)};
\draw[thick,XanaduRed] (-2.5,0) -- (-1.5,0);
\foreach \x/\l in {-2.5/$E_\text{lo}$,-1.5/$E_\text{hi}$, 1.0/$E_\text{max}$}
    \draw (\x,0.1) -- (\x,-0.1) node[below] {\l};
\draw (4,0) circle (1cm);
\draw[thick,XanaduPurple] (5,0) arc (0:180:1cm);
\draw[thick,XanaduGreen] (3,0) arc (180:360:1cm);
\draw[thick,XanaduRed] (3,0) arc (180:240:1cm);
\node at (5,1) {$\ket{0}$};
\node at (5,-1) {$\ket{1}$};
\node at (3.0, 1.2) {\scriptsize \shortstack{$E_k - E_{lo} <  0$}};
\node at (3.0, -1.2) {\scriptsize \shortstack{$E_k - E_{lo} > 0$}};
\draw[dashed] (2.75,0) -- (5.25,0);
\end{tikzpicture}
\caption{Assume $\bm{D}\ket{E_0}$ has support over the energy interval $[E_{\min}, E_{\max}]$, and we aim to project it into a narrower therapeutic energy window $[E_\text{lo}, E_\text{hi}]$. This can be achieved using a threshold projection approach: one projection from the left into $[E_{\min}, E_\text{hi}]$, and another from the right into $[E_\text{lo}, E_{\max}]$. In the Trotter setting (see \cref{app:TP_w_trotter}, the arc segments (blue and orange for (a), green and purple for (b)) are padded to equal lengths for symmetry. In qubitization, no padding is needed, the arcs are defined directly via the $\arccos(E/\lambda)$ map.
With this construction, a single qubit measurement suffices to determine whether the state lies inside or outside the target energy window. This approach relies on an appropriate shifting of the Hamiltonian to position the threshold at 0.}
\label{fig:median_lemma_variants}
\end{figure}

\begin{figure}[t]
\centerline{
\Qcircuit @C=1.2em @R=1.2em {
\lstick{\ket{0}} & \qw & \qw & \qw & \qw & \targ & \meter \\
\lstick{\ket{0}_\text{L}} & \qw & \qw & \gate{\text{In}} \ar @{-} [dd] & \qw & \ctrl{-1} & \qw \\
\lstick{\ket{0}_\text{R}} & \qw & \qw & \qw & \gate{\text{In}} \ar @{-} [d] & \ctrlo{-2} & \qw \\
\lstick{\ket{0}_\text{s}} &{/} \qw & \gate{\text{PREP}} & \gate{\text{QSP}} & \gate{\text{QSP}} & \gate{\text{PREP}^\dagger} & \meterB{\ket{0}}\\
}
}
\caption{High-level threshold projection circuit for determining whether an excitation energy lies within the therapeutic window $[E_\text{lo},E_\text{hi}]$. The system register $\ket{\psi}=\hat D\ket{E_0}$ is prepared via PREP and entangled with single-ancilla projectors,
which implement the conditions $E \leq E_\text{hi}$ (upper boundary) and $E \geq E_\text{lo}$ (lower boundary). Each ancilla undergoes a one-bit sign test using a sequence of controlled and uncontrolled walk operators.
A Toffoli gate writes the logical AND of the left and right projection ancillas, $\ket{0}_\text{L}$ and $\ket{0}_\text{R}$, into a joint ancilla initialized to $\ket{0}$. This ancilla is then measured in the computational basis; the outcome is $1$ if and only if both tests pass (in-window), and $0$ otherwise.
Additionally, afterward we apply PREP$^\dagger$ to the QSP output state and projected into $\ket{0}$, yielding a second independent measurement~\cite{loaiza2025simulating}. Together, these two readouts provide a $2$-fold reduction in sampling cost, as shown in~\cref{eq:outer_repetition_2}.
A detailed implementation of QSP is presented in~\cref{fig:QSP_double_phase_walk}.
}
\label{fig:algo_circs_dual_projection}
\end{figure}
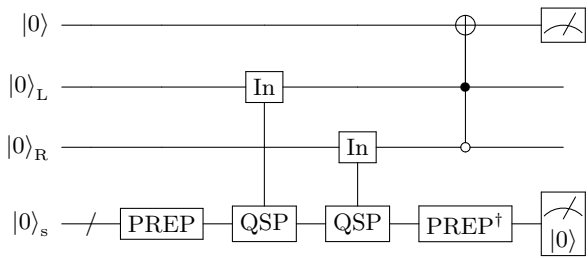

\begin{figure*}[t]
\centering
\[
\Qcircuit @C=1.2em @R=1.2em {
\lstick{\ket{0}_{L/R}} &  \qw& \gate{R(\theta_0,\phi_0, 0)} &  \ctrlo{1}  & \qw & \ctrlo{1} & \gate{R(\theta_1,\phi_1, 0)} &  \ctrlo{1}  & \qw & \ctrlo{1} & \qw & \cdots & &  \ctrlo{1}  & \qw & \ctrlo{1} & \gate{R(\theta_d,\phi_d, 0)} &\qw \\
\lstick{\ket{\psi}} & {/} \qw & \qw & \gate{R} & \gate{W} & \gate{R} & \qw & \gate{R} & \gate{W} & \gate{R}& \qw & \cdots &  & \gate{R} & \gate{W} & \gate{R}& \qw &  \qw \\
}
\]
\caption{The breakdown of the generalized QSP circuit used in \cref{fig:algo_circs_dual_projection} to evaluate the matrix element~\cref{eq:cumulative_absorption_sum}. The bulk of the circuit represents the implementation via quantum signal processing of a threshold function that probabilistically projects into the therapeutic window. The threshold function comprises the product of two Heaviside functions of the form $1/2 + \text{sign}(H-E_{\text{th}})/2$. To reduce the cost as much as possible, we leverage generalized quantum signal processing~\cite{motlagh2024generalized}.
As a further optimization, the application of $W$ or $W^\dagger$ are implemented sandwitching each $W$ by controlled reflections $\mathcal{R}$, see \cref{eq:walk_identity}~\cite{babbush2018encoding}. This can be combined with generalized QSP to halve the cost of QSP~\cite{berry2024doubling}. 
}
\label{fig:QSP_double_phase_walk}
\end{figure*}
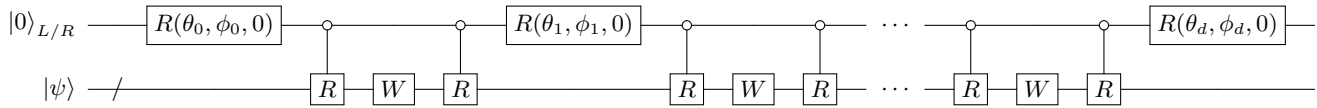

\begin{figure}[t]
    \centering
    \includegraphics[width=1.0\linewidth]{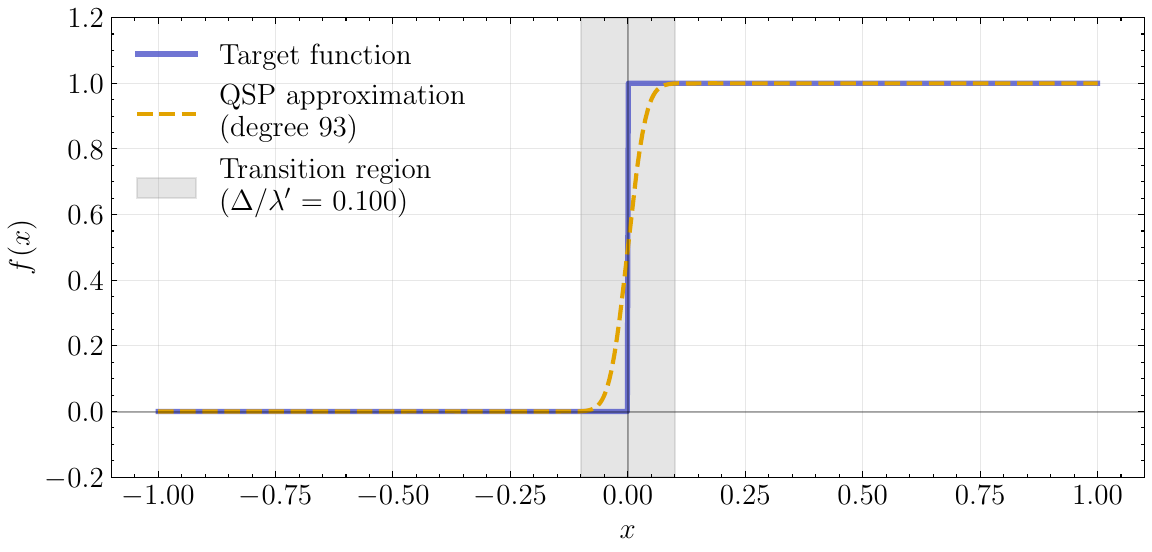}
    \caption{Quantum signal processing polynomial approximating a Heaviside function. While we could use a threshold function to synthesize the projector, using the software package \texttt{pyqsp} we found it more expensive trying to fit a single narrow threshold compared to two step functions: we chose the latter option~\cite{mrtc_unification_21,cdghs_finding_qsp_angles_20, haah_decomposition_19, gslw_qsvt_19, dmwl_efficient_phases_21}.}
    \label{fig:Heaviside}
\end{figure}

\textbf{Threshold projection:} 
The unit circle mapping used in our threshold projection algorithm is geometrically illustrated in \cref{fig:median_lemma_variants}, which shows how each projector encodes an energy threshold as a boundary on the unit circle. We define the spectral range $[E_\text{min}, E_\text{max}]$ of the system Hamiltonian. 
Within this range, we aim to isolate a subinterval $[E_\text{lo}, E_\text{hi}]$ corresponding to the therapeutic window. 
Our algorithm synthesizes a threshold function to probabilistically project this energy window subinterval. We use the software package \texttt{pyqsp}~\cite{mrtc_unification_21,cdghs_finding_qsp_angles_20, haah_decomposition_19, gslw_qsvt_19, dmwl_efficient_phases_21} to find polynomial approximations to this filter. In practice, we find it cheaper to implement two Heaviside functions rather than a single narrow threshold to accomplish this, see caption of~\cref{fig:Heaviside}.
The left projector tests whether the excitation energy satisfies $E_k \leq E_\text{hi}$, by targeting the interval $[E_\text{min}, E_\text{hi}]$ (shown in blue in \cref{fig:median_lemma_variants}). The right projector checks whether $E_k \geq E_\text{lo}$, by targeting $[E_\text{lo}, E_\text{max}]$ (shown in green/purple in \cref{fig:median_lemma_variants}). 
Each projection interval is mapped to a phase region symmetric about the equator (the $\ket{0}$–$\ket{1}$ decision boundary), allowing a single-qubit readout to cleanly distinguish whether an eigenstate lies above (below) the reference threshold $E_{\mathrm{lo}}$ ($E_{\mathrm{hi}}$).
Operationally, both projectors reduce to the same core task:  determining the sign of $E_k - E_{\mathrm{th}}$ for a given eigenvalue $E_k$. This is accomplished by simulating a shifted Hamiltonian, $H' = H - E_{\mathrm{th}}$, and extracting the sign of the expectation value $\langle E_k | H' | E_k \rangle = E_k - E_{\mathrm{th}}$. The sign directly determines the classification outcome. A high-level quantum circuit implementing this threshold projection is shown in~\cref{fig:algo_circs_dual_projection}. 

To map this decision into a binary ancilla measurement, the Hamiltonian is further rescaled to a dimensionless form, normalizing the relevant energy range to span the phase interval $(-\pi, \pi]$ on the unit circle in~\cref{fig:median_lemma_variants}. This ensures consistent encoding of energy thresholds as quantum phases. In our approach, this rescaling is implemented by the block encoding procedure detailed below. We also studied an alternative Trotter-based time evolution implementation, which turned out to be more expensive and thus less attractive for this application: the analysis and comparative resource estimates are provided in Appendix~\ref{app:TP_w_trotter}.

\textbf{Projectors:}  
We now describe the implementation of the quantum signal processing projector using qubitization, with the Hamiltonian block-encoded in its tensor hypercontraction (THC) form. 
An alternative implementation path is to use Trotter product formulas to implement phase estimation. However, achieving sufficiently small error---such that perturbations to eigenstates and eigenvalues remain within the spectral buffer around each decision threshold---requires a large number of Trotter steps, resulting in high Toffoli gate depth. Our analysis of this approach is given in~\cref{app:TP_w_trotter}.

In standard qubitization, the Hamiltonian is expressed in the Linear Combination of Unitaries (LCU) form: \begin{equation} H = \sum_{\ell=1}^{L} \alpha_\ell U_\ell, \quad \lambda = \sum_\ell |\alpha_\ell|, \end{equation} where $U_\ell$ are efficiently implementable unitaries and $\alpha_\ell$ are real coefficients, with $\lambda = \sum_\ell |\alpha_\ell|$ denoting the 1-norm of the Hamiltonian. The goal of our projection algorithm is to identify the sign of the shifted eigenvalue $E_k' = E_k - E_{\mathrm{th}}$, where $E_{\mathrm{th}}$ is the relevant threshold energy ($E_{\mathrm{lo}}$ or $E_{\mathrm{hi}}$).
To this end, we construct an LCU representation of the shifted Hamiltonian: 
\begin{equation}
\label{eq:LCU_shifted}
H' = \sum_{\ell=1}^{L} \alpha_\ell U_\ell - E_{\mathrm{th}} \cdot \mathbf{1}, \qquad \lambda' = \lambda + |E_{\mathrm{th}}|.
\end{equation} 
This shifted form is then block-encoded using standard primitives: a \textsc{PREPARE} circuit that loads the amplitudes ${\alpha_\ell}$, and a \textsc{SELECT} circuit that applies $U_\ell$ conditioned on an index register.
The resulting quantum walk operator encodes $H'/\lambda'$ into the eigenphase: 
\begin{equation} \label{eq:TP_walk_operator}
W = e^{\pm i \arccos(H'/\lambda')}. \end{equation} 

Next, we can use this walk operator to craft projectors that flag which energy range we are observing. We use generalized quantum signal processing~\cite{motlagh2024generalized} to implement two Heaviside functions that filter out the range of energy above $E_{\text{hi}}$ and below $E_{\text{lo}}$. These Heaviside functions are approximated with polynomials. Implementing a degree $d$ polynomial requires $d$ calls to the quantum walk operator $W$. The degree of such polynomial will depend on the width of the transition area. If we denote such \textit{transition width} by $\Delta/\lambda'$, the cost will be $\tilde{O}(\lambda'/\Delta)$. The corresponding generalized QSP circuit is shown in ~\cref{fig:QSP_double_phase_walk}. 

To lower the implementation cost, we exploit the double-phase trick introduced by Babbush et al.~\cite{babbush2018encoding}, which reduces the number of controlled walk operators required during phase estimation. 
In standard qubitization-based quantum phase estimation, powers of the walk operator $W$ must be applied conditionally on the phase register. However, directly controlling $W$ could be costly.
To mitigate this, we exploit the fact that the qubitized walk operator is defined as
\begin{equation}
    W = \mathcal{R} \cdot \textsc{PREP} \cdot\textsc{SEL}\cdot \textsc{PREP}^{\dagger}
\end{equation}
where $\mathcal{R} = (I-2|0\rangle\langle 0|) \otimes I$ is the reflection on the $\textsc{PREP}$ auxiliary qubits. 
This definition enables the transformation 
\begin{multline}
    \mathcal{R} (W)^k \mathcal{R}  = \mathcal{R} ( \mathcal{R} \cdot \textsc{PREP} \cdot\textsc{SEL}\cdot \textsc{PREP}^\dagger)^k \mathcal{R} \\
    = (  \textsc{PREP} \cdot\textsc{SEL}\cdot \textsc{PREP}^\dagger \cdot \mathcal{R})^k = (W^\dagger)^k.
\label{eq:walk_identity}
\end{multline}

This construction implements a symmetric phase-kickback:
\begin{equation}
|0\rangle|\psi\rangle \mapsto |0\rangle e^{+i E'_k t} |\psi\rangle,\quad
|1\rangle|\psi\rangle \mapsto |1\rangle e^{-i E'_k t} |\psi\rangle,
\end{equation}
effectively doubling the accumulated phase difference between eigenstates.
This symmetry is the basis of a factor-of-2 saving in quantum signal processing~\cite{berry2024doubling}.

\textbf{Tensor Hypercontraction:}
To reduce the implementation cost, tensor hypercontraction (THC) is used to approximately factorize the two-body part of the electronic Hamiltonian:
\begin{equation}
v_{pqrs} \approx \sum_{\mu, \nu=1}^M X_{p\mu} X_{q\mu} Z_{\mu\nu} X_{r\nu} X_{s\nu},
\end{equation}
where $X$ and $Z$ are real-valued tensors, and $M$ is the THC rank. 
The matrix $Z$ is symmetric, encoding the interaction strengths between auxiliary indices. The tensor $X$ defines the non-orthogonal basis rotation through a set of real-valued column vectors, which are normalized but not mutually orthogonal.
The rank $M$ is chosen to balance accuracy and compression. Specifically, we select $M$ such that the difference in Frobenius norm between the original two-electron integral tensor and its THC approximation 
is below a prescribed error threshold. 
In practice, this typically yields $M = \mathcal{O}(N)$, resulting in $L = \mathcal{O}(M^2)$ effective terms in the LCU expansion~\cite{lee2021_THC}. 

Another advantage of THC is to reduce the one-norm of the Hamiltonian, which can be further decreased with symmetry shifts~\cite{loaiza2023block,caesura2025}. Similarly, removing the identity term from the THC Hamiltonian---to reduce the one-norm of the Hamiltonian before we add the factor $E_{\text{th}}\bm{1}$---entails a global energy shift and thus a re-definition of the energies $E_{\min}$, $E_{\max}$, $E_{\text{lo}}$, $E_{\text{hi}}$.

In the qubitization framework, the SELECT oracle is implemented as a single, dynamically controlled circuit, driven by two auxiliary index registers, $\ket{\mu}$ and $\ket{\nu}$, each requiring $\lceil \log_2 M \rceil$ qubits. For a given basis state of these registers, the oracle uses QROM to load a corresponding set of precomputed rotation angles. These angles are used to synthesize the basis-change unitaries $U_\mu$ and $U_\nu$, which act on the system register, thereby replacing the need to directly control on the orbital indices. The orbital structure is therefore preserved but accessed indirectly through these dynamically constructed transformations. The algorithm's efficiency stems from the LCU expansion over only $\mathcal{O}(M^2)$ index pairs, significantly reducing control register size and SELECT complexity relative to full-rank decompositions.

While qubitization provides significant reductions in Toffoli implementation cost, it often requires a large number of logical qubits to store intermediate angle data, scratch registers, and ancilla address lines. 
To reduce ancilla cost, we adopt a width–depth trade-off inspired by the recent analysis of Caesura et al.~\cite{caesura2025}: load only $B$ rotation angles at a time into a shared $w$-bit register, apply the corresponding subset of Givens rotations that synthesize the basis-change unitaries $U_\mu$ and $U_\nu$ acting on the system register, and reuse the same register for subsequent batches. 

Let $L_\theta$ be the total number of angles per SELECT. By batching the rotations, the scratch register cost is reduced from $L_\theta w$ to $Bw$, where $w$ is the bit width per angle. As a consequence, the number of QROM calls increases from $1$ to $\lceil L_\theta/B\rceil$, trading qubits for depth. The extremes are $B=1$ (minimal width, maximal depth) and $B=L_\theta$ (maximal width, minimal depth). 
This qubit-gate trade-off approach allows us to tune the logical resource requirements of the SELECT circuit to best fit the characteristics of a given fault-tolerant quantum device.

\textbf{Results Sampling:}  
We evaluate whether an excitation from the dipole-weighted distribution $|\langle E_f | \hat{D} | E_0 \rangle|^2$ lies within the therapeutic window by performing both projections and counting the state only if both succeed. 
This process yields a binary indicator specifying whether the sampled excitation energy falls within the interval $[E_\text{lo}, E_\text{hi}]$. 
For the normalised dipole state of \cref{eq:normalized_state}, the probability of an ``in-window''
outcome,
\begin{align}
P_{\text{window}}
   &=
   \frac{1}{\mathcal{N}_D}
   \sum_{\Delta E_f \in [E_\text{lo}, E_\text{hi}]}
        |\langle E_f | \hat{D} | E_0 \rangle|^2 = \frac{A_{\text{window}}}{\mathcal{N}_D} 
\label{eq:p_window}
\end{align}
is, up to the known normalization factor $\mathcal N_D$ in~\cref{eq:normalization_Nd}, exactly the
\textit{cumulative dipole-weighted absorption} within the therapeutic
window.  

To estimate $P_{\text{window}}$, we then perform a \textit{sampling repetition}: we prepare fresh states and repeating the measurement procedure over $S$ shots. This produces a sequence of independent 0/1 outcomes, whose sample average is an unbiased estimator of $P_{\text{window}}$, and hence, after multiplying by $\mathcal N_D$, of the cumulative absorption in the window:
\begin{equation}
\label{eq:bernoulli_p}
    \widehat P_{\text{window}}
   \;=\;
   \frac{1}{S}\sum_{j=1}^{S} b_j,
   \qquad
   b_j \in \{0,1\}.
\end{equation}

In practice, we employ a \textit{double measurement}~\cite{loaiza2025simulating} strategy: each threshold projection circuit shot in \cref{fig:algo_circs_dual_projection} naturally yields two Bernoulli samples. The first is obtained by measuring the ancilla register, which directly signals whether the state lies inside the spectral window. The second is obtained from the overlap of the system register with the initial state after un-preparation. In other words, if the first measurement indicates we projected into the target energy window, the second measurement will sample from a binomial distribution with probability
\begin{multline}
    |(\braket{0|\text{PREP}^\dagger) (\Pi_{\Delta E_f\in[E_{\text{lo}}, E_{\text{hi}}]}\hat{D}|E_0})|^2 \\
    = \frac{1}{\mathcal{N}_D}\sum_{\Delta E_f\in[E_{\text{lo}}, E_{\text{hi}}]}|\braket{E_0|\hat{D}^\dagger |E_f}|^2 = \frac{A_{\text{window}}}{\mathcal{N}_D}.
\end{multline}
Similarly, if the first measurement indicated projection outside of the target energy window, the second will sample with a binomial distribution with probability $1-\frac{A_{\text{window}}}{\mathcal{N}_D}$. Consequently, this method provides another independent sample of the same probability distribution. Both outcomes are independent and identically distributed with mean $\hat{P}_{\text{window}}$, effectively doubling the amount of statistical data collected per shot.

\begin{figure}[t]
    \centering
    \includegraphics[width=1.0\linewidth]{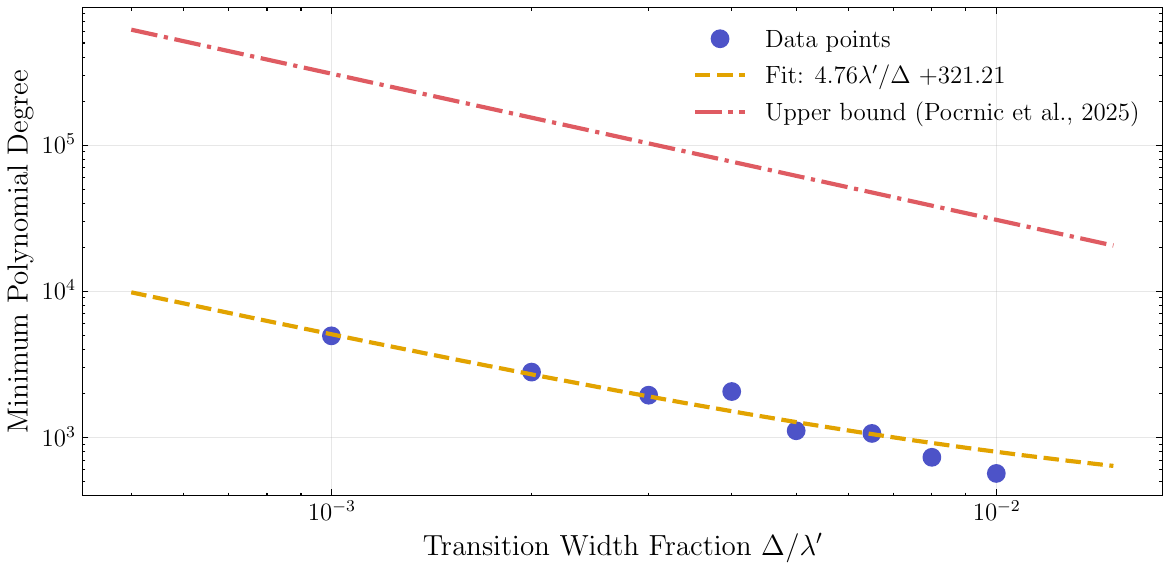}
    \caption{The polynomial degree required to approximate the Heaviside function to target error $\epsilon_{\text{H}} = 0.01$ was determined as a function of the transition width $\Delta/\lambda'$. The minimal-order polynomial was obtained using \texttt{pyqsp}~\cite{mrtc_unification_21, cdghs_finding_qsp_angles_20, haah_decomposition_19, gslw_qsvt_19, dmwl_efficient_phases_21}, via a binary search over degrees. A theoretical upper bound for the degree is given in Theorem 17 of  Ref.~\cite{pocrnic2025constant}.
    }
    \label{fig:degree_fit}
\end{figure}

\textbf{Cost analysis:} The threshold projection circuit cost is determined by two key components. The first is the cost of a single circuit shot, which depends on the spectral filtering sharpness and the fit accuracy outside it. The transition width $\Delta$ of the QSP filter---normalized by $\lambda'$---and the target precision $\epsilon_\mathrm{H}$ together set the minimum energy separation that can be resolved, and hence the required polynomial degree and walk cost.
The second component is the number of shots $S$ needed to achieve a desired statistical confidence, determined by the sampling error $\epsilon_\mathrm{samp}$ and failure probability $\delta_\mathrm{samp}$. In what follows, we break down the cost of these two components in terms of the following parameters:

\begin{enumerate}
    \item Circuit implementation cost: Each circuit implements 2 QSP projectors, as indicated in~\cref{fig:QSP_double_phase_walk,fig:algo_circs_dual_projection}. The factors determining the cost are:
\begin{itemize}
    \item Walk operator cost:
    The cost of the walk operator $W$ is determined by the required degree $d$, of the QSP polynomial. 
    While the polynomial fit in~\cref{fig:degree_fit} is a classical problem, the cost of implementing it as an energy filter in QSP is governed by the principles of quantum metrology. To resolve energies with a precision set by the filter's transition width $\Delta / \lambda'$, algorithms like QSP can operate at the Heisenberg limit. 
    This dictates that the degree $d$ of this polynomial grows as the inverse of the filter's transition width and polylogarithmically with the filter's precision, $\epsilon_{\mathrm{H}}$~\cite{Gily_n_2019}, according to the scaling relation:
    \begin{equation}
    \label{eq:qsp_degree_d}
        d = O\left(\frac{\lambda'}{\Delta} \text{poly}\log(\epsilon_{\mathrm{H}}^{-1})\right).
    \end{equation}
    Here, $\epsilon_{\mathrm{H}}$ represents the maximum allowable error in the filter's output, which is the amount the polynomial's $y$-value can deviate from the ideal 0 or 1. This algorithmic error can introduce a small systematic bias into the final estimated probability. To prevent this, we choose a small enough $\epsilon_{\mathrm{H}}$ to ensure the filter is highly accurate. The extra cost associated with this is minor compared to the additional statistical sampling cost we would incur from the filter's systematic bias: setting a tighter $\epsilon_\mathrm{H}$ is thus cheaper than repeating the entire algorithm many more times to estimate the final probability with a target precision $\epsilon_{\mathrm{samp}}$ in~\cref{eq:outer_repetition_1}.
    The $d$ calls to the walk operator $W$ (or $W^\dagger$) are implemented via controlled rotations as indicated in~\cref{eq:walk_identity}.
    
    \item Coefficient and rotation precision for Hamiltonian block encoding: although qubitization formally implements $e^{\pm \arccos(H'/\lambda')}$ exactly, the single-qubit rotations appearing in the quantum circuit must be specified to finite precision. In practice, the resource cost scales with the number of bits used to encode coefficients ($\aleph$) and load rotation angles ($\beth$):
    \begin{align}
    \label{eq:coeff_rot_precision}
        \aleph &= \left\lceil 2.5 + \log_2\left(\frac{\lambda'}{\epsilon_\text{coeff}}\right) \right\rceil, \\
        \beth &= \left\lceil 5.625 + \log_2\left(\frac{2\lambda' N_\text{orb}}{\epsilon_\text{rot}}\right) \right\rceil.
    \end{align}
    In this work, we choose $\epsilon_\text{coeff} = \epsilon_\text{rot} = 0.16 \text{ mHa}$~\cite{lee2021_THC, von_Burg_2021, bellonzi2024}. Note that $\beth$ is distinct from the phase-rotation precision used in the phase-gradient technique~\cite{lee2021_THC}, which we set equal to $\aleph$ in our analysis. 
\end{itemize}
If a single application of $W$ 
costs $G(\aleph, \beth)$ Toffoli gates---collecting the $\aleph$- and $\beth$-dependent costs of the \textsc{PREPARE} and \textsc{SELECT} oracles, and minor reflection/control overheads---the combined per-projection gate cost is:
\begin{equation}
\label{eq:proj_cost}
    C_\text{proj} (\epsilon_\mathrm{H},\frac{\Delta}{\lambda'}, \aleph, \beth) = d(\epsilon_\mathrm{H},\frac{\Delta}{\lambda'}) \cdot G(\aleph, \beth).
\end{equation}
In this work, we evaluate $G(\aleph,\beth)$ using an analytic cost model for qubitization based on the BLISS–THC~\cite{caesura2025} factorization, which yields a block-encoding of $H'/\lambda'$. The model accounts for the QROM-based implementations of \textsc{PREPARE} and \textsc{SELECT}, the bit-precision requirements for coefficients and rotations, and the qubit-depth trade-off introduced by parallel rotation. \texttt{PennyLane}'s resource estimator~\cite{bergholm2018pennylane} is used to obtain primitive gate counts. 

\item The sampling repetitions $S$ refer to the number of full threshold projection circuit evaluations (shots) needed to estimate $\hat{P}_{\mathrm{window}}$. Each circuit samples a Bernoulli random variable, equal to $1$ if the outcome lies inside the spectral window and $0$ otherwise. These variables are independent across shots and bounded in $[0,1]$, so the empirical mean $\hat{P}_\text{window}$ concentrates around the true probability $P_\text{window}$. By Hoeffding’s inequality, the number of samples required to bound the additive sampling error by $\epsilon_\mathrm{samp}$ with failure probability $\delta_\mathrm{samp}$ is
\begin{equation}
\label{eq:outer_repetition_1}
    S = \left\lceil \frac{1}{2\,\epsilon_{\mathrm{samp}}^{2}}\,
    \log\!\Big(\frac{2}{\delta_{\mathrm{samp}}}\Big)\right\rceil ,
\end{equation}
which guarantees $|\hat P_{\mathrm{window}}-P_{\mathrm{window}}|\le \epsilon_{\mathrm{samp}}$ with confidence at least $1-\delta_{\mathrm{samp}}$.

With the implementation of the \textit{double measurement} scheme, each circuit execution yields two statistically independent samples. This results in a factor-of-two reduction in the required number of physical shots, so we define the effective shot count as
\begin{equation}
\label{eq:outer_repetition_2}
    S_{\mathrm{dm}} = \tfrac{1}{2}S.
\end{equation}
\end{enumerate}

The overall cost of the threshold projection algorithm is obtained by combining the per-shot cost with the required number of sampling repetitions:
\begin{align}
\label{eq:TP_cost}
    n_\text{qubit} &= 2N + \max(n_{\text{SoS}}, n_\text{aux}+ 3)  ,\\
    C_\text{TP} &= S_\mathrm{dm} \cdot \left[ C_\text{SoS} +  2 \cdot C_\text{proj} (\epsilon_\mathrm{H},\Delta/\lambda', \aleph, \beth)\right],
\end{align}
where $n_\text{aux}$ denote the auxiliary qubits required to encode a single qubitised walk operator for the system. The state preparation cost $n_\text{SoS}(D) = 5\log_{2}D - 3$ and $C_\text{SoS}(D) = (2\log_{2}D - 2)D + 2^{\log_{2}D + 1} + D$ denotes the number of auxiliary qubits and Toffoli gates given in Ref.~\cite{Fomichev2024_SoS}, respectively. The $n_\text{SoS}$ auxiliary qubits can be recycled in the qubitization circuit, which leads to a peak ancilla count $\max(n_{\text{SoS}}, n_\text{aux}+ 3)$. The extra 3 qubits are the 3 additional auxiliary qubits in~\cref{fig:algo_circs_dual_projection}.

With the implementation described above, the threshold projection algorithm enables us to sample the normalized cumulative absorption $A_{\mathrm{window}}$ directly from the circuit illustrated in \cref{fig:QSP_double_phase_walk}.

\subsection{Evolution proxy algorithm for intersystem crossing rates}
\label{ssec:ISC_evo_proxy_algo}

A second key metric in evaluating photosensitizer performance is the efficiency of intersystem crossing from the excited singlet state to the triplet manifold. This spin-forbidden process underpins the generation of reactive triplet oxygen species and thus directly impacts the efficacy of photodynamic treatment. 
Accurate prediction of ISC rates requires accessing many-electron excited states and treating spin–orbit coupling (SOC) with high fidelity. However, wavefunction-based methods capable of resolving these effects are computationally prohibitive for the large, strongly correlated systems typically encountered in practice.

To overcome this challenge, we leverage the evolution-proxy quantum algorithm first developed in Ref.~\cite{ODMR_toyota_paper} to estimate ISC efficiency without requiring explicit rate calculations. The method probes the short-time dynamics induced by SOC, and computes a proxy observable that captures the leading-order singlet–triplet mixing ~\cite{ODMR_toyota_paper}:
\begin{equation}
\label{eq:ISC_proxy_matrix}
    \tilde k_{\text{ISC}}(t) = \left| \braket{E_{1,S=1, M} | e^{-it H_\SOC }|E_{1,S=0}} \right|^2.
\end{equation}
Here $E_{1}$ is the first excited state, $S$ is the spin quantum number, and $M$ the spin projection. In the short-time limit, this quantity grows linearly with $ t $ and is proportional to the SOC matrix element between the initial singlet and final triplet states. Specifically, a Taylor expansion yields:
\begin{multline}
    \left| \braket{E_{1,S=1, M} | e^{-it H_\SOC }|E_{1,S=0}} \right| \approx \\-it \cdot \left| \braket{E_{1, S=1, M} | H_{\SOC}|E_{1,S=0}} \right| + \mathcal{O}(t^3),
\end{multline}
which implies that the proxy quantity $\tilde k_{\text{ISC}}(t)$ is proportional to the exact ISC rate amplitude in the short-time limit:
\begin{equation}
    k_{\text{ISC}} \propto \left| \left\langle E_f \middle| H_{\text{SOC}} \middle| E_i \right\rangle \right|^2 \approx  \frac{\tilde k_{\text{ISC}}(t)}{t^2}.
\end{equation}
Therefore, the proxy rate $\tilde k_{\text{ISC}}$ serves as a useful quantity for assessing ISC efficiency, specifically in relative comparisons across candidate molecules.

While $\tilde k_{\text{ISC}}(t)$ is not an absolute ISC rate, it can still be used to assess whether a given molecular candidate has better or worse ISC efficiency through comparison to a reference system where ISC efficiency is known experimentally. Because the short-time expansion is applied consistently across all systems considered,
the computed values enable chemically meaningful \textit{ranking} of candidate compounds for ISC efficiencies.
This proxy forms the basis of our ISC screening strategy: we calibrate against a reference molecule with a known ISC rate and interpret $\tilde k_{\text{ISC}}(t)$ as a \emph{relative} ISC efficiency. A higher proxy value indicates stronger SOC-induced singlet–triplet mixing and thus improved ISC efficiency relative to the reference. This approach allows us to identify promising photosensitizer candidates without computing exact rates.

\begin{figure}[t]
\[
    \Qcircuit @C=1em @R=.7em {
  \ket{0} &&\qw & \multigate{1}{\text{SoS}} & \gate{S^\dagger} & \ctrl{1} &  \gate{H} &\meter \gategroup{1}{5}{1}{5}{.7em}{--}  \\
  \ket{0}_s &&{/}\qw & \ghost{\text{SoS}} & \multigate{1}{\text{QSP}}  & \gate{e^{-it H_{\SOC}}}  &\qw &\qw&  \\
  \ket{0} && \qw & \qw & \ghost{\text{QSP}} & \meter & &   &  & \\ 
    }
\]
    \caption{Modified Hadamard test circuit~\cite{ODMR_toyota_paper}. The initial state is prepared using the sum-of-Slaters method~\cite{Fomichev2024_SoS}, followed by single-ancilla coarse projection and postselection, similar to that described in \cref{ssec:TP_algo}. The presence (or absence) of the $S^\dagger$ gate selects the imaginary (or real) part of the matrix element in~\cref{eq:ISC_proxy_matrix}.}
    \label{fig:modified_Hadamard}
\end{figure}
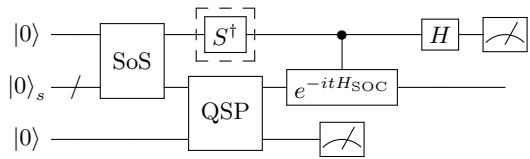

The quantum circuit representing the ISC evolution-proxy algorithm is presented in \cref{fig:modified_Hadamard}. We decompose the overall circuit into three stages: initial state preparation, short-time evolution under the SOC Hamiltonian, and measurement via a modified Hadamard test.
We now describe each component of the circuit in turn.

\textbf{State Preparation:} 
The first step in our approach is the preparation of initial states with support on the singlet and triplet manifolds. Classically, one of the main challenges in evaluating ISC rates lies precisely in state preparation, specifically, isolating states that participate in the ISC process. In our quantum algorithm, we begin by preparing a dipole-acted superposition of all optically active excited states using the sum-of-Slaters method~\cite{Fomichev2024_SoS}.

These initial states, however, typically have unwanted contributions from high-energy components. To suppress this contamination, we apply a projection step that probabilistically projects the state into a low-energy subspace, such as $[E_0, E']$. This procedure resembles the threshold projection scheme discussed in \cref{ssec:TP_algo}, but here it is applied in a `single-sided' manner~\cite{ODMR_toyota_paper} to filter out contributions above a specified energy threshold $E'$. The filtering circuit mirrors the threshold projection structure shown in \cref{fig:QSP_double_phase_walk}, relying on a small ancilla register together with qubitization. 
In the single-sided projection, only one energy cutoff is imposed, with the projector discarding amplitudes above the threshold and preserving  all low-lying excited states.

\textbf{Time evolution by $H_\SOC$:} Following state preparation, we perform time evolution under the SOC Hamiltonian, $H_{\text{SOC}}$, which captures the interaction between electron spin and orbital motion and drives intersystem crossing. 
Since $H_{\text{SOC}}$ is a one-body operator, time evolution under it can be fast-forwarded.
We apply a basis transformation to bring $ H_{\text{SOC}} $ into diagonal form, alleviating the need for a block encoding:
\begin{equation}
  h_{p\sigma,q\tau}^{\text{soc}} = U_0 Z_0 U_0^\dagger, \quad Z_0 = \mathrm{diag}(\lambda_1, \dots, \lambda_{2N}),
\end{equation}
where $U_0$ is the single-particle basis transformation matrix and $\{\lambda_{p\sigma}\}$ are the eigenvalues of $H_{\SOC}^{S, M}$.
This basis change induces a many-body unitary $\bm{U}_0$ (that can be synthesized using Thouless's theorem and decomposed using Givens rotations \cite{fomichev2025affordable}) that rotates the Fock-space state accordingly.
The full time-evolution operator is then a product of these unitaries and single-particle Pauli $Z$ rotations:
\begin{equation}
    e^{it H_\SOC^{S,M}} = \bm{U}_0 \prod_{p\sigma} e^{it \lambda_{p\sigma} \sigma_{z,p\sigma}} \bm{U}_0^\dagger,
\end{equation}
where $\sigma_{p\sigma}$ are Pauli $Z$ operators for orbitals $(p,\sigma)$. 

\textbf{Results Measurement:} Finally, to estimate the singlet–triplet transition amplitude, we employ a modified Hadamard test~\cite{parrish2019quantum,ODMR_toyota_paper}. 
In the standard Hadamard test, the ancilla would control our unitary $U(t)=e^{-itH_{\SOC}}$ on a single input state $\ket{\psi}$, so the measured ancilla expectation values would yield the diagonal element $\braket{\psi|U(t)|\psi}$. 
The modification presented in Refs.~\cite{parrish2019quantum,ODMR_toyota_paper} and implemented in this work consists of preparing a superposition of two different reference states in the ancilla branches,  
\begin{equation}
\label{eq:modified_hadamard_state}
    \ket{\zeta_0} \;=\; \frac{1}{\sqrt{2\gamma}} \left( \alpha \ket{0}\ket{\psi_{S=1,M}} + \beta \ket{1}\ket{\psi_{S=0}} \right),
\end{equation}
where $\ket{\psi_{S=1,M}}$ and $\ket{\psi_{S=0}}$ denote the prepared reference states chosen from the triplet ($S=1,M$) and singlet ($S=0$) sectors, respectively. Interference between the singlet and triplet components then gives direct access to the real and imaginary off-diagonal matrix element,
\begin{align}
\label{eq:evo_proxy_real_imagine}
\langle X \rangle_{\text{anc}} \;=\;  \text{Re} \left( \frac{\alpha^*\beta}{\gamma}\,\braket{\psi_{S=1,M} \mid U(t) \mid \psi_{S=0}} \right), \\
\langle Y \rangle_{\text{anc}} \;=\; \text{Im} \left( \frac{\alpha^*\beta}{\gamma}\,\braket{\psi_{S=1,M} \mid U(t) \mid \psi_{S=0}} \right),
\end{align}
This simple modification allows us to measure off-diagonal transition amplitudes directly, avoiding the need to square small overlaps.

\textbf{Cost Analysis:} We analyze the logical resource requirements of the ISC proxy algorithm based on the structure of its quantum circuit. 
The logical qubit requirement of the ISC proxy algorithm is modest: for an active space of $N$ orbitals, we require $2N$ spin-orbital qubits, and a register of size $n_{\text{SoS}}$ for the sum-of-Slaters state preparation. The Hadamard control qubit in \cref{fig:modified_Hadamard} can be absorbed into this ancilla register, while an additional qubit is needed for the QSP projector, and $n_\text{aux}$ auxiliary qubits are needed for walk operator encoding. Overall, the total logical qubit requirement is:
\begin{equation}
    n_\text{qubit} = 2N + \max(n_{\text{SoS}}, n_\text{aux}+ 2).
\end{equation}

The dominant Toffoli gate complexity arises from three components combined in the following equation, which we unpack in the following paragraphs:
\begin{align}
\label{eq:cost_evolution_proxy}
   & C_{\text{ev-pr}} = 4 S_{\mathrm{Had}}(\epsilon) \\
   & \left[ \frac{C_{\text{SoS}}(2D)+ C_\text{proj} (\epsilon_\mathrm{H},\Delta/\lambda', \aleph, \beth)}{\gamma^2} + C_{H_{\SOC}}\right].\nonumber
\end{align}
First, the cost of preparing multiconfigurational initial states with $D$ Slater determinants using the sum-of-Slaters method is $C_{\text{SoS}}(2D)$, same as described in~\cref{eq:TP_cost}.
Second, to suppress high-energy components in the prepared state, we apply a single-sided threshold projection. The cost of a single projection step, $C_{\text{proj}}$, is analyzed in detail in \cref{ssec:TP_algo},~\cref{eq:proj_cost}. 
Third, the short-time evolution operator $ e^{-it H_{\SOC}} $ is fast-forwardable due to the one-body structure of $ H_{\SOC} $, and its implementation cost is denoted by $C_{H_{\SOC}}$, which scales as $O(N^2)$. 
This routine is run for both the real and imaginary components of the matrix element in \cref{eq:evo_proxy_real_imagine} (a factor of two) and for two separable spin projections ($M=0$ and $M=\pm1$, another factor of two), resulting in the overall prefactor of $4$.
Finally, the modified Hadamard test used to estimate the ISC proxy observable requires $ S_{\text{Had}}(\epsilon) $ repetitions, defined in Ref.~\cite{ODMR_toyota_paper}, to reach a target precision $ \epsilon$ in the proxy measurement. Here, $\gamma$ is a normalization factor from~\cref{eq:modified_hadamard_state}, and $\gamma^2 $ denotes the success probability of the postselection step within the Hadamard test circuit.

With the implementation described above, the ISC proxy algorithm offers a resource-efficient surrogate for the exact ISC rate, enabling us to extract the effective intersystem crossing rate $\tilde k_{\text{ISC}}$ directly from short-time dynamics of the electronic register.
In regimes where spin–orbit coupling is weak and vibronic effects dominate, however, a more explicit treatment of vibrational degrees of freedom might be required. To address such cases, in \cref{app:ISC_dynamics_algo} we give a high-level outline of a vibronic dynamics–based algorithm that accounts for vibrationally assisted nonradiative pathways competing with ISC, along with a corresponding resource analysis and approximate constant-factor estimates.

\section{Application: BODIPY Derivatives}
\label{sec:application}
Having presented the quantum algorithms to be used for designing better photosensitizers, in this section we apply them to realistic systems.
Among various classes of  photosensitizers, boron–dipyrromethene (BODIPY) derivatives have gained sustained attention due to their high molar absorption coefficients, narrow emission bands, and excellent photochemical and thermal stability in biological environments~\cite{Loudet2007, Ulrich2008}.
Their optical properties are highly tunable through structural modification, enabling access to the therapeutic window, $700\text{--}850$\,nm, with minimal background absorption. 
While strategies such as halogenation, donor–acceptor substitution, or incorporation of transition metals are known to enhance spin–orbit coupling and promote triplet formation \cite{Wang2020, Sunny2023}, their impact on absorption and ISC is not straightforward to predict. This motivates simulation, which can systematically evaluate how specific substitutions affect both properties.
Water-soluble and biocompatible derivatives have been developed that preserve their absorption profiles in biological environments while exhibiting minimal side effects in the absence of light~\cite{Kamkaew2013, Li2020}. These features make BODIPY derivatives promising candidates for photodynamic cancer therapy and related biomedical applications.

Our concrete system selection reflects two guiding principles: first, to include photosensitizer candidates that are of active interest in the PDT and molecular photophysics communities; and second, to span a range of chemical modifications and orbital sizes, thereby demonstrating the versatility and scalability of our quantum simulation approach.

Following our discussion of instance selection and active space design in \cref{ssec:bdp_active_space} below, we present instance- and size-specific resource estimates to demonstrate the practical feasibility of our algorithm. The results indicate that accurate simulations of key properties of BODIPY photosensitizers are achievable using a few hundred logical qubits and Toffoli gate depths ranging from about $10^7$ for the smallest active spaces to roughly $10^9$ for the most demanding cases examined in this work.

\subsection{BODIPY candidates}
\label{ssec:bdp_active_space}

Our set of demonstrative instances begins with the parent BODIPY scaffold, orig-BDP. This compact and chemically stable core has well-defined spectroscopic properties and serves as the baseline for evaluating the other BODIPY derivatives in this study.

Building on this, we examine Br-BDP, a brominated aza-BODIPY developed by Gallagher and O’Shea et al.~\cite{Byrne2009}, which demonstrated promising preclinical photodynamic efficacy and was highlighted as having potential for future Phase~I clinical evaluation. The incorporation of bromine and phenyl-extended aza units enhances intersystem crossing (ISC), boosting ROS generation via heavy-atom effects.

As a heavy-atom-free alternative, we include a triazolyl-substituted aza-BODIPY (trz-aza-BDP) recently synthesized and experimentally characterized by Hlogyik et al~\cite{Hlogyik2023}. This compound features strong NIR absorption near $654$\,nm, negligible toxicity in a dark environment, and induces significant cell damage upon light activation. Its ability to achieve high ROS yields without relying on halogens or metals highlights the role of charge-transfer character and acceptor substitution in promoting ISC. The absence of bromine or iodine also addresses concerns related to environmental impact and synthetic accessibility.

Finally, we consider Pt-BDP, a platinum–BODIPY complex reported by Bera et al.~\cite{Bera2022_BODIPY_Pt}. Coordination to a transition metal enhances ISC via spin–orbit coupling while introducing strong d-orbital correlation effects. 
These multireference features and the associated large active-space demands make the system particularly challenging for conventional electronic structure methods, and thus a compelling target for quantum simulation.

Together, these four systems provide a diverse yet chemically coherent set for quantum resource estimation. They collectively span a broad range of design strategies explored in photosensitizer development. 

\begin{figure*}[t]
    \centering
    \includegraphics[width=0.85\linewidth]{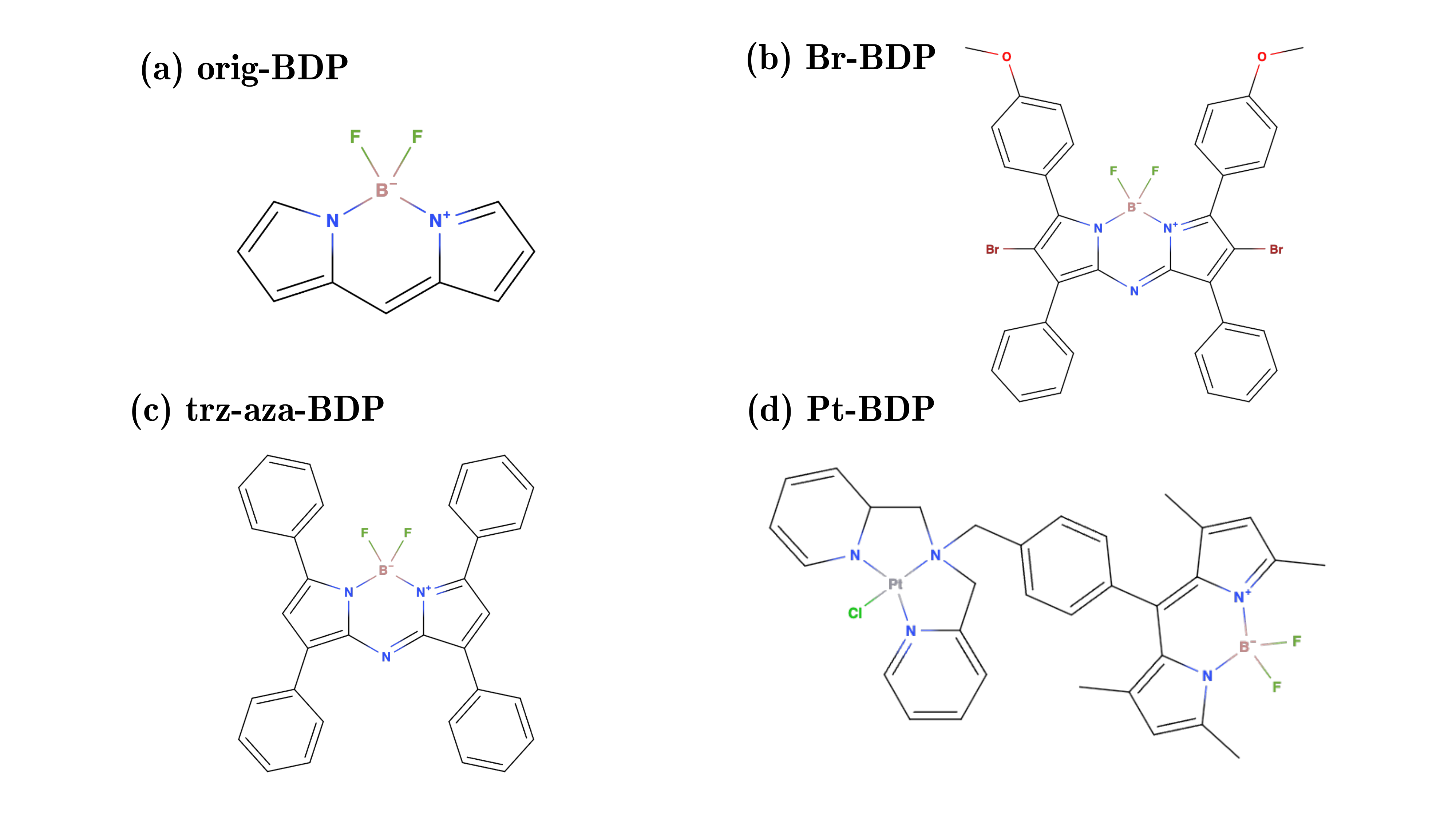}
    \caption{Molecular structures of the four BODIPY derivatives studied in this work. Each structure reflects substituent modifications that tune the electronic and photophysical properties relevant to PDT.}
    \label{fig:bodipy_structures}
\end{figure*}

Active space methods are standard in quantum chemistry for focusing computational effort on the orbitals most relevant for correlation and spectroscopy. Instead of treating the full orbital space---which is often unnecessary for capturing the physics of interest---we isolate chemically meaningful subsets that preserve accuracy while controlling computational complexity. This same philosophy carries over to quantum simulation: active spaces enable accurate modeling of strongly correlated states while keeping resource requirements feasible.
To study how quantum resource requirements scale with model fidelity, we define three levels of active spaces for each BODIPY instance: small (S), medium (M), and large (L). These are constructed to span a range of electronic complexity, from minimal qualitative descriptions to more extended treatments that incorporate additional chemically relevant orbitals and approach quantitative accuracy. 

We begin with the basic BODIPY framework, orig-BDP, whose active space definition is guided by the benchmark study of Momeni and Brown~\cite{Momeni_2015}. In that work, the authors systematically increased the active space size from the minimal CAS(2,2)---containing only the highest occupied molecular orbital (HOMO) and lowest unoccupied molecular orbital (LUMO)---to a larger CAS(12,11) that includes all $\pi$ electrons of the conjugated system as well as the nitrogen lone pair.
CASPT2 calculations with the latter active space yield vertical excitation energies in close agreement with experiment, indicating that this level of treatment is sufficient for this relatively small system. 
We adopt this configuration as a chemically validated starting point. However, our objective is not to re-compute orig-BDP's spectrum, but to use it as a well-characterized reference for validating our quantum algorithms. Moreover, while such an active space remains tractable classically for orig-BDP, scaling to larger systems quickly renders high-level classical methods infeasible. 

To systematically explore the scaling of quantum resource demands with system complexity, we extend this baseline in a chemically informed manner across all molecules studied. Active spaces are constructed to reflect photophysically relevant electronic features:
the $\pi$-conjugated orbitals of the BODIPY core together with the nitrogen lone pairs---corresponding to the full CAS[12,11] of the Momeni–Brown reference---are always retained as the minimal photophysically relevant set;
additional orbitals are incorporated based on the presence of structural modifications such as halogens, transition metals and triazole groups, which impact electronic structure and enhance spin-orbit coupling. Phenyl $\pi$-systems and high-lying virtual orbitals are selectively included in the medium and large active spaces. 
This leads to three tiers of active spaces constructed by incrementally including orbitals of decreasing chemical and electronic relevance to the target photophysics.
Such a design ensures chemically faithful representations of clinically relevant BODIPY derivatives while also providing a systematic way to track how quantum resource demands evolve for increasingly complex, real-world systems.
The resulting orbital counts are summarized in Table~\ref{tab:active_space}.

\begin{table}[h!]
\centering
\renewcommand{\arraystretch}{1.5}
\setlength{\tabcolsep}{10pt}
\begin{tabular}{lccc}
        \multicolumn{1}{c}{} & \multicolumn{3}{c}{Active Space Size}\\ %
        \cline{2-4}
Instances & Small & Medium & Large \\
\hline
\hline
Orig-BDP  & 11  & 15   & 19  \\
Trz-aza-BODIPY & 11  & 19  & 35  \\
Br-aza-BDP   & 17  & 21  &  45   \\
Pt-BDP & 16 & 24 & 30 \\

\hline
\end{tabular}
\caption{Active space sizes (number of spatial orbitals) for each BODIPY derivative at small, medium, and large levels. Active space selection is based on chemical relevance: core $\pi$-conjugated orbitals and nitrogen lone pairs are consistently included~\cite{Momeni_2015}; triazole, halogen, and transition metal orbitals are added based on their contributions to excitation and spin–orbit effects; medium and large spaces further include extended $\pi$-systems and high-lying virtual orbitals to capture delocalization and correlation effects.}
\label{tab:active_space}
\end{table}

The quantum simulations performed in this work rely on molecular Hamiltonians constructed over chemically motivated active spaces, as described in \cref{ssec:bdp_active_space}. 
For each BODIPY instance, we generate the effective electronic Hamiltonian $H_\text{eff}$ from mean-field-level orbitals and integrals using PySCF~\cite{libcint2015, pyscf2018, pyscf2020}, The two-body part of $H_\text{eff}$ was decomposed into its low-rank factorized form using OpenFermion~\cite{openfermion2019}. This representation is subsequently mapped into a form suitable for quantum time evolution using qubitization, as described in \cref{ssec:TP_algo}.

This low-rank factorized construction also naturally supports the inclusion of static environmental effects. 
In particular, dielectric screening from solvent environments can be introduced with a frequency-independent polarizable continuum model (PCM), which adds a reaction-field potential obtained from surface polarization charges on the molecular cavity to the one- and two-electron integrals:
\begin{align}
t_{pq}^{\mathrm{solv}} &= 
  t_{pq}^{\mathrm{eff}}
  + \int\!\!\int
      \chi_p^{*}(\mathbf{r})\,
      \frac{\sigma(\mathbf{s})}{\lvert \mathbf{r}-\mathbf{s}\rvert}\,
      \chi_q(\mathbf{r})\,
      d\mathbf{r}\,d\mathbf{s}, \\[4pt]
v_{pqrs}^{\mathrm{solv}} &= 
\varepsilon^{-1}\,v_{pqrs}^{\mathrm{eff}},
\end{align}
$\varepsilon$ is the static dielectric constant of water. 

Importantly, this correction can be incorporated without requiring additional self-consistency at the post-SCF level. Theoretically, since the PCM solvent field is a classical response to the SCF-level charge density, it can be treated as an external one-body potential that remains fixed during subsequent correlation treatments or Hamiltonian transformations. For weakly polar solutes like BODIPY derivatives, this frozen-field approximation is particularly well-controlled.
This approach is also supported by prior benchmarks in correlated electronic structure theory. For example, Ref.~\cite{nishimoto2025} reports that keeping the solvent reaction field fixed at the SCF level during CASPT2 yields excitation energies within chemical accuracy of those obtained with a fully relaxed solvent field. Iterating the solvent response during post-SCF treatments incurs significant computational overhead while offering limited benefit. As a result, one-shot inclusion of solvent-modified integrals is adopted for the applications in this work.

For BODIPY excitation energies in the therapeutic range, where water exhibits negligible electronic absorption, such treatment is valid within the non-resonant regime~\cite{wildman_2019, liu_2019}. The solvent-corrected integrals can then be processed identically through the THC pipeline, adding no extra quantum-circuit overhead.

\subsection{Resource Estimates}
\label{ssec:RE_results}

Using the approaches presented in \cref{sec:algos}, we now obtain concrete resource estimates for computing the cumulative absorption and intersystem crossing rate for the set of BODIPY derivatives described in \cref{ssec:bdp_active_space}. 
We use the THC form of the Hamiltonian as described in \cref{ssec:TP_algo}, and estimate the number of logical qubits and Toffoli gates required to simulate the key observables for each BODIPY instance.
The active spaces used for each system are summarized in \cref{tab:active_space}. For both the threshold projection and evolution proxy algorithms, we assume a sum-of-Slaters
with $10^4$ determinants is used 
for preparing the initial state~\cite{Fomichev2024_SoS}.

\textbf{Cumulative Absorption with threshold projection:}
The walk operator $W$ in \cref{eq:TP_walk_operator} is constructed with coefficient and rotation precision $\aleph$ and $\beth$. both tuned to a cost per walk $G(\aleph, \beth)$ in~\cref{eq:proj_cost}.
Each threshold projection circuit shot involves two projections, each implemented with a QSP polynomial of degree $d = O(\lambda'/\Delta\text{poly}\log {\epsilon}_{\mathrm{H}}^{-1})$. 
The sharpness of the QSP filter is governed by the target precision $\epsilon_\mathrm{H}$ and the transition width $\Delta/\lambda'$. 
Here, $\epsilon_{\mathrm{H}}$ denotes a dimensionless additive error in approximating the sign function, which quantifies the residual amplitude leakage between in-window and out-of-window eigenstates.
To obtain concrete resource estimates, we empirically fitted the polynomial degree $d$ required to approximate the Heaviside function for ${\epsilon}_{\mathrm{H}} \approx 0.01 \ll \epsilon_{\mathrm{samp}}$, as shown in \cref{fig:degree_fit}. 
The best fit is
\begin{equation}
    d = 4.7571 \frac{\lambda'}{\Delta} +  321.2051,
\end{equation}
with $R^2 = 0.969733$, where this value indicates how much of the data’s variation is captured by the fit (values close to 1 mean a strong fit). For comparison, we also display the upper bound from Theorem 17 of Ref. ~\cite{pocrnic2025constant}.

Together, the walk cost $G(\aleph, \beth)$ and required polynomial degree $d$ determine the resources for a single circuit execution. 
At the next level, one must account for sampling statistics. Estimating the cumulative absorption probability$ \hat{P}_\text{window}$ in \cref{eq:p_window} requires repeating the full procedure $S$ times. With a target additive sampling error $\epsilon_\mathrm{samp}$ corresponding to a $10\%$ resolution in the estimated cumulative absorption probability $\hat{P}_\text{window}$ and a sampling failure probability of $\delta_\mathrm{samp} = 0.01$, we obtain $S = 1.32 \times 10^{2}$. 
Using qubitization and a width–depth trade-off strategy described in \cref{ssec:TP_algo}, we set each SELECT block to load only $B = 1$ rotation angle per QROM call. This minimizes ancilla overhead at the cost of increased QROM depth, requiring $\lceil L_\theta / B \rceil$ queries, where $L_\theta$ is the total number of rotation angles.

The resulting resource estimates $C_\text{TP}$ with the above set of parameters are reported in \cref{tab:RE_results}.
Despite aggressively trading qubits for depth, the total Toffoli count required remains reasonable. 
For the largest system considered (45 spatial orbitals, 90 spin orbitals), each threshold projection circuit requires $2.48 \times 10^9$ Toffoli gates with $346$ logical qubits. This represents the high end of the resource range we estimate, whereas all other systems studied in this work fall between $10^7$ and $10^8$ Toffoli gates, well within the capabilities expected realistic fault-tolerant devices.

\textbf{ISC with Evolution Proxy Algorithm:}
In the evolution proxy circuit, we project into the low-energy singlet and triplet subspaces using the same method as in threshold projection, but with projection applied only once rather than on both sides. As a result, the single projection cost remains $C_\text{proj}$ as above.
To evaluate singlet–triplet mixing, we perform fast-forwarded controlled time evolution under the one-body $H_\text{SOC}$. 
In the modified Hadamard test, we target an observable accuracy of $0.1$. Given a projection success probability of $\gamma = 0.7$ for each state (see \cref{eq:modified_hadamard_state}), this leads to $S_\text{Had} = 483$ total samples.

With all parameters specified to evaluate the ISC proxy rate cost $C_\text{ev-pr}$, the resulting resource estimates across different active space sizes are reported in \cref{tab:RE_results}.
The per-shot Toffoli count is generally in the $10^7$–$10^8$ range and rises to $\sim 10^9$ for the large size active space considered in~\cref{tab:active_space}, remaining comparable in scale to threshold projection but with nearly twice the circuit size.
This overhead arises despite using a single projector, owing to the additional cost of simulating $H_\text{SOC}$ and the multiplicative factor of four introduced in \cref{eq:cost_evolution_proxy}.

\section{Conclusions}
\label{sec:conclusions}
This work shows how quantum computing can be used in drug design of photodynamic cancer therapy agents. We focus on these agents because their complex electronic interactions are difficult for classical computers to model accurately.
By enabling direct access to key photophysical properties, quantum algorithms offer a systematic route to accelerate photosensitizer discovery and reduce reliance on costly empirical screening.

Specifically, 
we have developed a fault-tolerant quantum computational framework for predicting two core photophysical observables critical to photosensitizer efficacy: cumulative absorption within the $700\text{--}850$\,nm therapeutic window, and intersystem crossing (ISC) rates. 
Cumulative absorption is targeted using a novel threshold projection algorithm; meanwhile ISC rates can be obtained via the short-time evolution-proxy algorithm, optionally supplemented with a vibronic dynamics simulation in the presence of strong internal conversion. 

When applied to clinically relevant BODIPY derivatives, including heavy-atom and metal-substituted variants, our proposed algorithms show favorable resource requirements across active spaces ranging from 11 to 45 spatial orbitals.
Threshold projection requires fewer than $350$ logical qubits and $10^7$–$10^9$ Toffoli gates, with the upper end corresponding to the largest system studied.
The ISC algorithms incur costs about twice those of threshold projection but remain in the same overall range.
For completeness, we also considered a vibronic dynamics–based ISC computation algorithm to cover cases where vibronic coupling drives nonradiative pathways. As detailed in \cref{app:ISC_dynamics_algo}, its estimated cost is somewhat higher, on the order of $\sim$$10^{10}$ Toffoli gates. In regimes where vibronic coupling is not the primary channel, the evolution-proxy algorithm provides the more practical route for estimating ISC rates. 
The combination of threshold projection and the evolution-proxy algorithm yields resource estimates
indicating that quantum simulations of photosensitizers at a scale and accuracy beyond the reach of classical methods are a realistic near-future prospect.

%

Looking ahead, there are several avenues for enhancing the capability and predictive power of our screening framework. Immediate progress can be made by lowering quantum resource requirements through algorithmic refinements, such as more efficient Hamiltonian simulation or optimized state preparation. 
A natural next step is to extend our present focus on absorption and ISC to a more complete description of the photochemical pathway. In hypoxic tumors, where oxygen availability is extremely low, Type~I electron-transfer processes dominate and are responsible for generating highly damaging radical species. The efficiency and molecular targets of these reactions are difficult to probe experimentally and are classically hard to model because they involve correlated charge transfer between the photosensitizer and nearby biomolecules. Capturing Type~I reactivity in such environments therefore represents a promising direction for future quantum simulations. 
Finally, beyond these algorithmic improvements, an important frontier is modeling photosensitizers within realistic biological microenvironments, where dielectric response, available acceptor states, and local structure can differ significantly from idealized models.

Overall, the optimized quantum algorithms described here and the attractive resource estimates obtained for clinically relevant systems suggest that quantum computing could play a prominent role in the discovery and optimization of next-generation photosensitizers for cancer therapy.

\bibliography{refs}

\section{Acknowledgments}

This research used resources of the National Energy Research Scientific Computing Center (NERSC), a Department of Energy User Facility using NERSC award DDR QIS-ERCAP ERCAP0032729.

\clearpage
\newpage
\appendix

\onecolumngrid

\section{Quantum Algorithms for ISC: vibronic dynamics approach}
\label{app:ISC_dynamics_algo}

While a purely electronic Hamiltonian approach provides a resource-efficient way to estimate ISC rates and is often adequate when spin-orbit coupling (SOC) between singlet and triplet states is sufficiently strong, it becomes insufficient in regimes where SOC is weak or vibronic effects play a significant role. 
In these situations, nonradiative pathways such as internal conversion (IC) can act as competing relaxation channels, rapidly depleting the excited singlet population before ISC has a chance to occur. IC refers to a spin-conserving, non-radiative transition between electronic states of the same multiplicity (e.g., $E_{n=1; S=0} \rightarrow E_{n=0; S=0}$), often facilitated by nuclear motion and vibrational coupling.

In such regimes, explicitly including vibrational modes becomes important to capture the system's dynamics accurately. Vibrational relaxation, shown schematically as step~\textcircled{4} in \cref{fig:type2_PDT_diagram}, refers to the ultrafast, nonradiative dissipation of excess vibrational energy within an electronic state. After photoexcitation, the molecule typically relaxes to the lowest vibrational level of the singlet excited state before undergoing further radiative or nonradiative transitions. This process shapes the initial conditions from which both IC and ISC proceed and therefore plays a key role in determining their relative competition.

To capture the effect of vibronic coupling and simulate ISC dynamics explicitly, we adopt the vibronic dynamics quantum algorithm introduced in Ref.~\cite{xanadu_vibronic}, which enables efficient simulation of the spin-vibronic Hamiltonian on a quantum computer. 
This approach models population transfer between electronic states of different spin multiplicities as a dynamical process governed by coupled nuclear and electronic degrees of freedom. 

The central observable of interest is the total triplet population $P_T(t)$, defined as the probability of finding the system in any triplet electronic state at time $t$. In our notation, the system wavefunction is $\ket{\psi(t)} \in \mathcal{H}_\text{el} \otimes \mathcal{H}_\text{vib}$, the full electronic–vibrational Hilbert space. A triplet state is labeled $\ket{E_{n;S=1;M}}$, where $S = 1$ denotes spin multiplicity and $M$ the spin projection.
The observable is expressed as:
\begin{equation}
P_T(t) = \sum_{n, M} \left\langle \psi(t) \middle| \left( \ket{E_{n;S=1;M}}\bra{E_{n;S=1;M}} \otimes \mathbb{I}_\text{vib} \right) \middle| \psi(t) \right\rangle,
\end{equation}
where $\mathbb{I}_\text{vib}$ is the identity operator on the vibrational register, ensuring that we trace over vibrational degrees of freedom while projecting onto the triplet electronic subspace.

In the weak spin-orbit coupling regime, $P_T(t)$ typically increases slowly from zero and may be well approximated by a single-exponential kinetic form $P_T(t) \approx 1 - \exp(-k_{\mathrm{ISC}} t)$ in the early-time limit~\cite{Gotardo2017}. The ISC rate constant is then extracted from the initial slope:
\begin{equation}
\label{eq:first_order_ISC_rate}
k_{\mathrm{ISC;vib}} \approx \left. \frac{dP_T(t)}{dt} \right|_{t \to 0}.
\end{equation}
This approach provides a dynamics-based definition of $k_{\mathrm{ISC;vib}}$ that captures the full effect of spin and vibrational couplings without relying on perturbative assumptions.

\textbf{Spin-Vibronic Hamiltonian:} The system is described by the Koppel–Domcke–Cederbaum (KDC) Hamiltonian, which couples vibrational, electronic and spin degrees of freedom:
\begin{equation}
H = \mathbb{I}_{\mathrm{el}} \otimes (T_{\mathrm{nuc}} + V_0) + W'(Q).
\label{eq:vibrarional_H}
\end{equation}
Here, $T_{\mathrm{nuc}}$ denotes the nuclear kinetic energy operator and $V_0$ is a reference harmonic potential energy surface (PES) centered at the equilibrium geometry. Nuclear motion is described relative to this baseline, with each electronic state corresponding to a displaced or distorted PES:
\begin{equation}
\label{eq:vibronic_T_and_V}
T_{\text{nuc}} = \frac{1}{2} \sum_r \omega_r P_r^2, \quad
V_0 = \frac{1}{2} \sum_r \omega_r Q_r^2.
\end{equation}
The diabatic potential $W'(Q)$ encodes nuclear-coordinate-dependent couplings between electronic states, incorporating both vibronic and spin–orbit interactions essential for spin-changing processes like ISC.
This diabatic coupling block is expanded as a Taylor series about the equilibrium geometry:
\begin{equation}
W'_{ij}(Q) = \lambda^{(ij)} + \sum_r a_r^{(ij)} Q_r + \sum_{r,s} b_{rs}^{(ij)} Q_r Q_s + \cdots,
\end{equation}
where $i$ and $j$ index the electronic diabatic states, including singlet and triplet configurations, and $Q_r$ denotes the normal mode displacement along vibrational mode $r$. The constant term $\lambda^{(ij)}$ captures purely electronic coupling at the equilibrium geometry, while $a_r^{(ij)}$ and $b_{rs}^{(ij)}$ represent linear and quadratic vibronic coupling coefficients, respectively. Higher-order terms may be retained as needed, but are truncated here under the assumption that linear and quadratic couplings capture the dominant vibronic effects near the equilibrium geometry.

\textbf{Quantum Simulation Procedure:} 
The dynamics are simulated by initializing the system in a separable product state $| \psi \rangle = | E_{n; S} \rangle \otimes_r^{M_{vib}-1} | \chi_0^{(r)} \rangle$, where $|E_{n; S}\rangle$ denotes an initial electronic eigenstate indexed by excitation level $n$ and spin quantum number $S \in {0,1}$ (singlet or triplet), and $|\chi_0^{(r)}\rangle$ is the vibrational ground state of mode $r$. The total number of modes $M_{vib}$ defines the vibrational Hilbert space dimension.
Each vibrational mode is discretized in real space, enabling efficient implementation of operators like $e^{i \theta Q_r}$ and $e^{i \theta P_r^2}$from \cref{eq:vibronic_T_and_V}. 

Time evolution under the spin–vibronic Hamiltonian $H$ (\cref{eq:vibrarional_H}) is performed using a second-order Trotter–Suzuki product formula:
\begin{equation}
e^{-iHt} \approx \left[\prod_j e^{-i H_j \Delta t/2} \prod_j e^{-i H_j \Delta t/2} \right]^{t/\Delta t},
\label{vibronic_trotter}
\end{equation}
where the Hamiltonian is decomposed as $H = \sum_m H_m$, with each $H_m$ constructed to allow efficient implementation via block-diagonalization in the electronic subspace. This fragmentation strategy, rather than treating $T_{\mathrm{nuc}}$, $V_0$, and $W'(Q)$ separately, partitions the Hamiltonian into fragments that each target a specific structure in the multistate diabatic coupling matrix. The nuclear kinetic energy term is treated as a separate fragment and exponentiated in the momentum basis using QFT, while each potential fragment $H_m$ is exponentiated in real space via fast-forwarded arithmetic operations.
The Trotter step size $\Delta t$ is selected to balance discretization error against circuit depth, with bounds derived from commutator norms ensuring the total simulation error remains below a chosen threshold~\cite{xanadu_vibronic}. 
The number of (spin-diabatic) electronic states $N$ included in the model is chosen based on the physical mechanism being probed.
To capture the early-time singlet–triplet population transfer, it is sufficient to include a small set of low-lying singlet and triplet states that are vibronically and spin–orbit coupled. Accordingly, we restrict the electronic register to the minimal subspace required to resolve the dominant ISC pathway. 
After evolution for a short time $t$, repeated measurements are performed on the electronic register to estimate the instantaneous triplet population $P_T(t)$. By analyzing the short-time behavior of $P_T(t)$, we extract the ISC rate constant $k_{\mathrm{ISC;vib}}$ from the initial slope in \cref{eq:first_order_ISC_rate}. 

Following the process described here, we can model photosensitizer behaviour during photoexcitation using a vibronic model. This simulation is complementary to the purely electronic ISC approach described in the previous section, as it allows to capture the competition between spin–flip transitions and vibrational relaxation---enabling simulation of ISC dynamics in the weak SOC regime where vibronic effects dominate.

\textbf{Cost Analysis}
The resource requirements of the vibronic dynamics ISC algorithm have been quantified in Ref.~\cite{xanadu_vibronic}. For a system with $N$ electronic states, $M$ vibrational modes, and $K$ grid points per mode, a single second-order Trotter step can be implemented using
\begin{equation}
O\!\big(N M^{d} \, (d \log^{2} K + N)\big)
\end{equation}
Toffoli gates and 
\begin{equation}
    O\!\big(d^{2} \log K + \log N\big)
\end{equation}
ancilla qubits, where $d$ is the polynomial degree of the diabatic expansion. The scaling reflects three key ingredients: block-diagonalization of each fragment reduces non-Clifford cost to polynomial in $Q_r$; QROM-based coefficient loading introduces an $O(N)$ overhead; and a caching scheme reuses intermediate products, lowering the arithmetic cost of higher-degree monomials. 
To simulate evolution for time $t$, this step must be repeated $r = t/\Delta t$ times, where $r$ is chosen to ensure the total Trotter error is below the target tolerance.

\textbf{Resource Estimation:}
To improve vibronic dynamics simulation efficiency, we do not simulate the full characteristic timescale $ \tau = 1 / k_{\mathrm{ISC}}$ to obtain accurate rate estimates. 
Instead, we follow a strategy similar to that of Northey and Penfold~\cite{northey_penfold_2018}, who inferred a complete ISC timescale of $125$~ns based on only $100$~ps of simulated quantum dynamics.
This shorter trajectory was sufficient to capture the early-stage population transfer and yielded a ISC rate $ k_{\mathrm{ISC}} = 8.0 \times 10^6~\mathrm{s}^{-1} $, in good agreement with experimental observations. 
Building on this precedent, we apply a similar reduction factor of approximately 1000.
In the case of BODIPY-based systems, ISC is expected to occur on a faster timescale than the Northey and Penfold application case. For example, Wang et al.\cite{Wang2020} reported a characteristic ISC timescale of $ \tau = 8~\mathrm{ns} $. 
We therefore simulate vibronic dynamics over an $8$~ps window, which we assume to be  sufficient to extract a reliable rate estimate.

For a quadratic vibronic coupling model involving $N = 5$ spin–vibronic states and $M = 19$ vibrational modes, and targeting a $1\%$ error tolerance, we estimate a total of $3.7 \times 10^5$ Trotter steps are required to perform time evolution for 8 ps. The corresponding quantum resource estimates are $146$ logical qubits and $3.1 \times 10^{10}$ Toffoli gates.

\section{Frequency-Dependent Solvent Effect}
\label{app:bosonic_solvent}
We propose a quantum embedding scheme that approximates solvent polarization effects through a set of harmonic oscillators (bosonic modes). Each bosonic mode corresponds to a pole in the solvent's dielectric response function $\varepsilon(\omega)$ and collectively reproduces the frequency-dependent polarization of the environment in the relevant spectral window.

This construction draws on the classical $\omega + \text{pole}$ model, where the solvent dielectric function is represented by a sum over damped oscillators. The imaginary part of the dielectric function defines a spectral density $J(\omega)$, which can be discretized into a set of effective oscillators:
\begin{equation}
J(\omega) \approx \sum_k \frac{\pi g_k^2}{2} \delta(\omega - \omega_k)
\end{equation}
where $\omega_k$ is the resonance frequency of mode $k$ and $g_k$ is its coupling strength. These parameters can be extracted by fitting experimental dielectric data of water using a Drude–Lorentz model, which ensures that the quantum embedding faithfully reflects the classical solvent dispersion.

In the $700\text{--}850$\,nm range relevant for BODIPY excitation, water has no resonant transitions. Therefore, a small number of modes (1-3) is sufficient to approximate the real part of $\varepsilon(\omega)$ across this range, while the imaginary part is negligible. 

The total Hamiltonian of the system is then given by:
\begin{equation}
H = H_{\text{elec}} + H_{\text{solv}} + H_{\text{int}}
\end{equation}
where:
\begin{align}
H_{\text{elec}} &= \sum_{pq} h_{pq} a_p^\dagger a_q + \frac{1}{2} \sum_{pqrs} (pq|rs) a_p^\dagger a_q^\dagger a_s a_r \\
H_{\text{solv}} &= \sum_k \hbar \omega_k \left(b_k^\dagger b_k + \frac{1}{2} \right) \\
H_{\text{int}} &= \sum_{k,pq} g_k^{(pq)} (b_k + b_k^\dagger) a_p^\dagger a_q
\end{align}
Here, $a_p^\dagger$ and $a_q$ are fermionic creation and annihilation operators acting on the solute orbitals, while $b_k^\dagger$ and $b_k$ are bosonic ladder operators for the solvent mode $k$. The interaction Hamiltonian couples the charge distribution of the solute (via one-electron density operators $a_p^\dagger a_q$) to the bosonic environment modes with strength $g_k^{(pq)}$.

In the non-resonant regime, the bosonic modes do not dynamically evolve in response to the solute’s excitation because the solute frequency $\omega_s$ lies far from any solvent resonance. Instead, the solvent undergoes a slight, static polarization shift in response to the solute’s electric field, with no time-dependent feedback. This allows the bosonic environment to be safely frozen in its ground state.

The resulting effective Hamiltonian for the solute includes a solvent-induced static perturbation to the one-electron part of the molecular Hamiltonian (w/ second order correction):
\begin{equation}
H_{\text{eff}} = H_{\text{elec}} + \sum_{pq} V_{pq}^{\text{solv}}\, a_p^\dagger a_q
\end{equation}
where the solvent-induced correction $V_{pq}^{\text{solv}}$ is obtained by integrating out the bosonic modes:
\begin{equation}
V_{pq}^{\text{solv}} = - \sum_k \frac{ (g_k^{(pq)})^2}{\hbar \omega_k}
\end{equation}

This correction captures the frequency-filtered polarization of the solvent and can be viewed as a quantum analog of the classical reaction field in polarizable continuum models. Crucially, since $V_{pq}^{\text{solv}}$ is precomputed and static, no real-time solvent dynamics or entanglement with the environment is required during the quantum simulation. This results in substantial reductions in quantum resource requirements while maintaining physically meaningful solvent effects.

\section{QPE and threshold projection with Trotterization}
\label{app:TP_w_trotter}
Similar to the qubitization algorithm described in the main body, we present a Hamiltonian simulation framework based on the Trotter product formula.
In threshold projection, we must ensure that each projection onto the boundaries of the therapeutic window can be evaluated using a single qubit. To achieve this, the spectral range is padded as necessary to create a symmetric mapping, and the Hamiltonian is shifted and rescaled into a dimensionless form for each projection:
\begin{equation}
\widetilde{H}_{\mathrm{L}} = \tau\,(H - E_\mathrm{hi}),
\qquad
\widetilde{H}_{\mathrm{R}} = \tau\,(H - E_\mathrm{lo}),
\label{eq:TP_rescaled_Hamiltonian}
\end{equation}
where 
$E_\mathrm{hi}$ and $E_\mathrm{lo}$ are reference energies used to center the spectrum in \cref{fig:median_lemma_variants}(a) and \cref{fig:median_lemma_variants} (b) correspondingly and 
$\tau$ is a scaling factor that normalizes the projection interval to span $[-\pi, \pi]$, ensuring consistent encoding of energy thresholds as quantum phases. 

To maintain symmetry in the wrapped phase representation, both projection intervals must span equal arc lengths on the unit circle. This may require artificial padding of the original spectral bounds by extending either $E_\text{min}$ or $E_\text{max}$ beyond the actual energy support of the system. 
For instance, in the case illustrated in \cref{fig:median_lemma_variants}(a), if $[E_\text{min}, E_\text{hi}]$ (blue segment) is shorter, we extend $E_\text{min}$ leftward to define an extended lower bound $\widetilde{E}_{\min}$; if the right interval (orange segment) is shorter, we extend $E_\text{max}$ rightward to balance the construction. The padded full spectral width is denoted as $\Lambda$, and we define the corresponding scaling factor $\tau$ for this projection to be:
\begin{equation} \label{eq:rescaling_tau}
\tau = \frac{2\pi}{\Lambda_{L/R}},
\end{equation}
where $\Lambda_L$ and $\Lambda_R$ refer to the padded full spectral width used for the upper and lower projectors, respectively.
This symmetric rescaling ensures that each coarse projector maps its respective interval to a clean half-circle, enabling accurate binary classification with minimal ancilla overhead.

To determine the time-evolution cost, we impose a constraint that limits the maximum allowable energy bias introduced by Trotterization. Specifically, we require the systematic energy shift to remain smaller than a fixed fractional width of a spectral bin:
\begin{equation}
\label{eq:half_bin}
|\Delta E|_{\mathrm{sys}} \leq \xi \delta E_{\mathrm{bin}},
\end{equation}
where $\delta E_{\mathrm{bin}} = E_{\mathrm{hi}} - E_{\mathrm{lo}}$ denotes the total energy width of a bin, and $0<\xi\ll1$ so that we tolerate a shift no larger than a fixed fraction of the full window.
This constraint ensures that Trotterization does not shift eigenvalues across the boundary of the target energy window, which would lead to incorrect evaluation of the cumulative absorption.

For efficient simulation, we generate the effective electronic Hamiltonian $H_\text{eff}$ from mean-field-level orbitals and integrals. These are subsequently transformed into a representation suitable for quantum time evolution using compressed double factorization (CDF)~\cite{fomichev2025affordable, cohn2021quantum}, which provides a compact encoding of both one- and two-body terms.
The one-electron integrals are diagonalized exactly,
\begin{equation}
t_{pq}^\text{eff} = \sum_k \tilde{U}^{(0)}_{pk} \tilde{Z}^{(0)}_{kk} \tilde{U}^{(0)}_{qk},
\end{equation}
while the two-electron part is approximated as a low-rank expansion,
\begin{equation}
v^\text{eff}_{pqrs} \approx \sum_{\ell=1}^{L} \sum_{k,l=1}^N U^{(\ell)}_{pk} U^{(\ell)}_{qk} Z^{(\ell)}_{kl} U^{(\ell)}_{rl} U^{(\ell)}_{sl}.
\end{equation}
This form rewrites the Hamiltonian as a sum of block-diagonal terms, each diagonal in a rotated single-particle basis defined by $U^{(\ell)}$. Using the Jordan-Wigner transform~\cite{Jordan1928}, the resulting operators can be directly mapped to Pauli strings dominated by number operators and pairwise $Z$ terms. The Hamiltonian takes the following form:
\begin{align}
    H &= \left(E + \sum_k Z_k^{(0)} -\frac{1}{2}\sum_{\ell,kl}Z_{kl}^{(\ell)}+\frac{1}{4} \sum_{\ell,k} Z^{(\ell)}_{kk} \right)\bm{1}\nonumber\\ \label{eq:CDF_Hamiltonian}
    &-\frac{1}{2}\bm{U}^{(0)} \left[\sum_k Z^{(0)}_{k} \sum_\gamma  \sigma_{z,k\gamma} \right](\bm{U}^{(0)})^T\\
    &+\frac{1}{8}\sum_\ell  \bm{U}^{(\ell)} \left[\sum_{(k,\gamma)\neq(l,\tau)}  \left( Z^{(\ell)}_{kl} \sigma_{z,k\gamma} \sigma_{z,l\tau}\right) \right](\bm{U}^{(\ell)})^T.
\end{align}

As an alternative to qubitization with THC, we implement the controlled evolution operator using the second-order symmetric Suzuki–Trotter product formula:
\begin{equation}
\label{eq:symmetric_trotter}
e^{-i\Delta \tilde{H}} \approx \prod_{j=1}^{M} e^{-i \frac{\Delta}{2} \tilde{H}_j} \prod_{j=M}^{1} e^{-i \frac{\Delta}{2} \tilde{H}_j},
\end{equation}
where $\tilde{H} = \sum_{j=1}^{M} \tilde{H}_j$ and the $\tilde{H}_j$ are additive components from \cref{eq:CDF_Hamiltonian}. 
For regular QPE with Trotter, 
the Trotter step size $ \Delta $ defines the evolution operator as: 
\begin{equation}
U_2(\tilde{H}, \Delta) = \exp\left[ -i \Delta \left( \tilde{H} + \Delta^2 \tilde{Y}_3 + \mathcal{O}(\Delta^4) \right) \right],
\end{equation}
where the leading error term is expressed as
\begin{equation}
\tilde{Y}_3 = \tau^3 Y_3,
\end{equation}
with $ Y_3 $ defined according to the Baker--Campbell--Hausdorff  expansion~\cite{BLANES2004135} as 
\begin{equation}\label{eq:Y3}
 Y_3:=\sum_{j} \Bigg[ \frac{1}{12}{[H_j,[ \sum_{h<j}H_h, H_j]]} + \frac{1}{24}{[\sum_{h<j} H_h,[ \sum_{h<j}H_h, H_j]]} \Bigg].
\end{equation}
Here, the $ H_j $ denote the additive components of the unscaled Hamiltonian $ H = \sum_j H_j $, and the ordering condition $ h < j $ reflects the symmetric Trotter splitting. This expression matches the standard form used in Trotter error analysis for second-order product formulas, and allows a perturbative interpretation of the simulation as exact time evolution under an approximate Hamiltonian.
This expression corresponds to exact evolution under the effective Hamiltonian
\begin{equation}
\tilde{H}_{\mathrm{eff}} = \tilde{H} + \Delta^2 \tau^3 Y_3.
\end{equation}
To quantify the systematic energy bias introduced by Trotterization, we consider the spectral norm of the deviation between $ H_{\mathrm{eff}} $ and $ \tilde{H} $,
\begin{equation}
\|\tilde{H}_{\mathrm{eff}} - \tilde{H}\| = \Delta^2 \tau^3 \|Y_3\| = c \tau^3 \Delta^2,
\end{equation}
where we define $ c \equiv \|Y_3\| $. Dividing by one factor of $ \tau $ to return to unscaled energy units yields
\begin{equation}
\label{eq:energy_shift}
|\Delta E|_{\mathrm{sys}} \leq c \tau^2 \Delta^2.
\end{equation}
This bound 
reflects a direct perturbative correction to energy levels. 
Combining Eq.~\eqref{eq:half_bin} with Eq.~\eqref{eq:energy_shift} yields a maximum allowable Trotter step size that preserves spectral resolution:
\begin{equation}
\label{eq:delta_max}
\Delta_{\max} = \sqrt{ \frac{\xi \delta E_{\mathrm{bin}}}{c \tau^2} }.
\end{equation}
This constraint ensures that the binning structure used to define the absorption observable is not blurred by numerical artifacts introduced during time evolution.

From a perturbative standpoint, the leading energy shift induced by the symmetric
Trotter formula is
\begin{equation}
\delta E \;=\; \tau^{2}\,\Delta^{2}\,\langle E|Y_{3}|E\rangle,
\end{equation}
where $|E\rangle$ and $E$ are eigenstates and eigenvalues of the physical Hamiltonian
$H$.  Off–diagonal state mixing appears at the same order, but our bias criterion
targets \emph{eigenvalue} drift, so we retain only the diagonal contribution.
Bounding the matrix element by the spectral norm
$c\equiv\|Y_{3}\|$ gives the worst-case systematic error in ~\cref{eq:energy_shift}. 

We use the bound in Eq.~\eqref{eq:delta_max} to set the Trotter time step.  
The level of tolerances
can be adjusted by substituting tighter or looser
$\xi$ in the formula.

For large active spaces, direct computation of $ \|Y_3\| $ is infeasible, so we construct a heuristic estimate. 
Assuming the two-electron Hamiltonian is represented in compressed double-factorized form
\begin{equation}
H = \sum_{k=0}^{M-1} Z^{(k)} \otimes Z^{(k)}, \qquad M = \mathcal{O}(N),
\end{equation}
with $ Z^{(k)} \in \mathbb{C}^{N \times N} $, we observe that each commutator between $ N \times N $ slices can generate at most $ \mathcal{O}(N^2) $ Pauli strings. 


If instead threshold projection is applied with Trotter, the padded spectral width $\Lambda$ sets the evolution time for the controlled Hamiltonian simulation in threshold projection to
$T_\text{TP} = 2\pi/\Lambda$. Finally, each QSP projector approximates a degree $d$ polynomial that costs $d$ calls to the walk operator to implement.
The total cost is given as:
\begin{align}\label{eq:cost_evolution_proxy}
   & C_{\text{TP;Trotter}} = K \cdot
   \left[ C_{\text{SOS}}(D)+ 2 \cdot d \cdot C_{\text{Trot}}(H_\text{eff}) \right].\nonumber
\end{align}

We report the complementary resource estimation results for threshold projection and ISC evolution-proxy algorithms all implemented with Trotter time-evolution.  

\begin{table*}[t]
\centering
\setlength{\tabcolsep}{9pt} 
\renewcommand{\arraystretch}{1.2}
    \begin{tabular}{c c c c c c c}
        \multicolumn{1}{c}{} & \multicolumn{3}{c}{QPE with Trotter} & \multicolumn{3}{c}{ISC evolution proxy with Trotter}\\ 
    \cline{2-7}
        $N$ & Qubits & Toffoli per shot  & Number of Shots $S_\mathrm{dm}$ & Qubits & Toffoli per shot & Number of Shots $S_\mathrm{Had}$ \\
        \hline
        \hline

    \multicolumn{7}{c}{\centering Original BODIPY} \\ \hline
     $11$ & $86$ & $1.22 \times 10^{9}$ & $1.32 \times 10^{2}$ & $91$ & $3.48 \times 10^{9}$ & $4.83 \times 10^{2}$\\
     $15$ & $94$ & $3.10 \times 10^{9}$ & $1.32 \times 10^{2}$ & $99$ & $8.87 \times 10^{9}$ & $4.83 \times 10^{2}$ \\
     $19$ & $102$ & $6.33 \times 10^{9}$ & $1.32 \times 10^{2}$ & $107$ & $1.81 \times 10^{10}$ & $4.83 \times 10^{2}$  \\

\hline \multicolumn{7}{c}{\centering Triazolyl aza-BODIPY (THC edited) } \\ \hline 
     $11$ & $86$ & $1.22 \times 10^{9}$ & $1.32 \times 10^{2}$ & $91$ & $3.48 \times 10^{9}$ & $4.83 \times 10^{2}$ \\
     $19$ & $102$ & $6.33 \times 10^{9}$ & $1.32 \times 10^{2}$ & $107$ & $1.81 \times 10^{10}$ & $4.83 \times 10^{2}$ \\
     $35$ & $134$ & $3.98 \times 10^{10}$ & $1.32 \times 10^{2}$ & $139$ & $1.14 \times 10^{11}$ & $4.83 \times 10^{2}$ \\
      
\hline \multicolumn{7}{c}{\centering Brominated aza-BODIPY (THC edited)} \\ \hline
     $17$ & $98$ & $4.53 \times 10^{9}$ & $1.32 \times 10^{2}$ & $103$ & $1.29 \times 10^{10}$ & $4.83 \times 10^{2}$  \\  
     $21$ & $106$ & $8.56 \times 10^{9}$ & $1.32 \times 10^{2}$ & $111$ & $2.45 \times 10^{10}$ & $4.83 \times 10^{2}$ \\
     $45$ & $154$ & $8.48 \times 10^{10}$ & $1.32 \times 10^{2}$ & $159$ & $2.42 \times 10^{11}$ & $4.83 \times 10^{2}$ \\

\hline \multicolumn{7}{c}{\centering Pt-BODIPY (THC edited)
} \\ \hline
     $16$ & $96$ & $3.77 \times 10^{9}$ & $1.32 \times 10^{2}$ & $101$ & $1.89 \times 10^{10}$ & $4.83 \times 10^{2}$ \\ 
     $24$ & $112$ & $1.28 \times 10^{10}$ & $1.32 \times 10^{2}$ & $117$ & $6.40 \times 10^{10}$ & $4.83 \times 10^{2}$ \\
     $30$ & $124$ & $2.50 \times 10^{10}$ & $1.32 \times 10^{2}$ & $129$ & $7.16 \times 10^{10}$ & $4.83 \times 10^{2}$ \\

        \hline
    \end{tabular}
    \caption{Logical resource estimates for two key spectroscopic quantities across active spaces of $N$ spatial orbitals. 
    Cumulative absorption is obtained using QSP with a Suzuki--Trotter Hamiltonian simulation, 
    while the ISC rate is evaluated using the evolution-proxy algorithm. 
    The ``QSP + Trotter'' costs assume second-order Trotterization with step number chosen such that the per-segment simulation error 
    is below the allocated polynomial approximation error budget. }
\label{tab:app_RE_results_trotter}
\end{table*}

\end{document}